\documentclass[pra,twocolumn,showkeys,showpacs,amsmath,amsfonts,floatfix,superscriptaddress,nofootinbib]{revtex4-2}

\usepackage{amsmath}
\usepackage{amssymb}
\usepackage{amsfonts,}
\usepackage{bm}
\usepackage[colorlinks=true,citecolor=blue,linkcolor=blue,urlcolor=blue]{hyperref}
\usepackage{graphicx}
\usepackage{color}
\usepackage[dvipsnames]{xcolor} 
\usepackage[utf8]{inputenc}
\usepackage{bbold}
\usepackage{diagbox}
\usepackage{tabularx}

\global\long\def\p{\prime}
\global\long\def\ket#1{|#1\rangle}
\global\long\def\bra#1{\langle#1|}
\global\long\def\proj#1#2{|#1\rangle\langle#2|}
\global\long\def\inner#1#2{\langle#1|#2\rangle}
\global\long\def\tr{\mathrm{tr}}
\global\long\def\im{\imath}

\newcommand{\dg} {{\dagger}}
\newcommand{\pd} {{\phantom\dagger}}
\newcommand{\f}[2] {{#1}_{#2}^{\pd}}
\newcommand{\fd}[2] {{#1}_{#2}^\dg}
\newcommand{\uf}[2] {{\underline{#1}}_{#2}^{\pd}}
\newcommand{\ufd}[2] {{\underline{#1}}_{#2}^\dg}

\newcommand{\ci}[1] {{{c}_{#1}^{\pd}}}
\newcommand{\cid}[1] {{c}_{#1}^\dg}
\newcommand{\bG} {\bm{G}}
\newcommand{\bS} {\bm{\Sigma}}
\newcommand{\bGa} {\bm{\Gamma}}
\newcommand{\bH}{\bm{\bar{H}}}
\newcommand{\bHS}{\bm{\bar{H}}_\qs}

\renewcommand{\Im} {\operatorname{Im}}
\newcommand{\ql}{\mathcal{L}}
\newcommand{\qs}{\mathcal{S}}
\newcommand{\qr}{\mathcal{R}}
\newcommand{\qi}{\mathcal{I}}
\newcommand{\qu}{\mathcal{U}}
\newcommand{\qw}{\mathcal{W}}
\newcommand{\ok}{\proj{v_k}{v_k}}

\newcommand{\kk}{\ket{v_k}}
\renewcommand{\[}{\begin{equation}}
\renewcommand{\]}{\end{equation}}
\newcommand{\action}{{\tilde \tau}}
\newcommand{\TS}{\tau_\qs}
\newcommand{\TW}{\tau_\qw}
\newcommand{\TT}{\tau_\mathrm{th}}
\newcommand{\timeNESS}{\tau_c}
\newcommand{\onS}{\omega_\qs}
\newcommand{\rhoI}[1]{\rho^{\mathrm{eq}}_{#1}}
\newcommand{\rhoo}{\rho^{\circ}_\qs}
\newcommand{\Io}{I^\circ}
\newcommand{\CSo}{\cm^\circ_\qs}
\newcommand{\qd}{\mathcal{D}}
\newcommand{\NW}{N_\qw}
\newcommand{\NS}{N_\qs}
\newcommand{\NT}{N_\mathrm{th}}
\newcommand{\OSEE}{S_O}
\DeclareMathAlphabet\mathbfcal{OMS}{cmsy}{b}{n}
\newcommand{\cm}{\mathbfcal{C}}
\newcommand{\ucm}{\underline{\cm}}
\newcommand{\Gop}{\mathbfcal{G}}
\newcommand{\Pop}{\mathbfcal{P}}
\newcommand{\CC}{C}
\newcommand{\sig}{\sigma_I} 
\newcommand{\asigsq}{\overline \sigma_I^2} 
\newcommand{\asig}{\overline \sigma_I} 

\begin{document}

\title{Accumulative reservoir construction: \\ Bridging continuously relaxed and periodically refreshed extended reservoirs}

\author{Gabriela W\'ojtowicz}
\affiliation{Jagiellonian University, Institute of Theoretical Physics, \L{}ojasiewicza 11, 30-348 Krak\'{o}w, Poland}
\affiliation{Biophysical and Biomedical Measurement Group, Microsystems and Nanotechnology Division, Physical Measurement Laboratory, National Institute of Standards and Technology, Gaithersburg, Maryland, USA}
\author{Archak Purkayastha}
\affiliation{School of Physics, Trinity College Dublin, College Green, Dublin 2, Ireland}
\affiliation{Centre for complex quantum systems, Aarhus University, Nordre Ringgade 1, 8000 Aarhus C, Denmark}
\author{Michael Zwolak}
\email{mpz@nist.gov}
\affiliation{Biophysical and Biomedical Measurement Group, Microsystems and Nanotechnology Division, Physical Measurement Laboratory, National Institute of Standards and Technology, Gaithersburg, Maryland, USA}
\author{Marek M. Rams}
\email{marek.rams@uj.edu.pl}
\affiliation{Jagiellonian University, Institute of Theoretical Physics, \L{}ojasiewicza 11, 30-348 Krak\'{o}w, Poland}

\date{\today}

\begin{abstract}
The simulation of open many--body quantum systems is challenging, requiring methods to both handle  exponentially large Hilbert spaces and represent the influence of (infinite) particle and energy reservoirs. These two requirements are at odds with each other: Larger collections of modes can increase the fidelity of the  reservoir representation but come at a substantial computational cost when included in numerical many--body techniques. An increasingly utilized and natural approach to control the growth of the  reservoir is to cast a finite set of reservoir modes themselves as an open quantum system. There are, though, many routes to do so. Here, we introduce an accumulative  reservoir construction---an ARC---that employs a series of partial refreshes of the extended  reservoirs. Through this series, the representation accumulates the character of an infinite reservoir. This provides a unified framework for both continuous (Lindblad) relaxation and a recently introduced periodically refresh approach (i.e., discrete resets of the reservoir modes to equilibrium). In the context of quantum transport, we show that the phase space for physical behavior separates into discrete and continuous relaxation regimes with the boundary between them set by natural, physical timescales. Both of these regimes ``turnover'' into regions of over-- and under--damped coherence in a way reminiscent of Kramers' crossover. We examine how the range of behavior impacts errors and the computational cost, including within tensor networks. These results provide the first comparison of distinct extended reservoir approaches, showing that they have different scaling of error versus cost (with a bridging ARC regime decaying fastest). Exploiting the enhanced scaling, though, will be challenging, as it comes with a substantial increase in (operator space) entanglement entropy.
\end{abstract}

\maketitle

\section{Introduction}

The interplay between the dynamics of isolated quantum systems and processes induced by macroscopic reservoirs, such as dissipation and decoherence, is central to the understanding and design of quantum devices. Yet, the resulting composite systems are challenging to simulate. Overcoming this challenge will lead to more than just accurate modeling of phenomena, such as transport and many--body relaxation. It will also provide new avenues for the experimental realization of open quantum systems---ones that exploit the control and stability of ultra-cold atom~\cite{lewenstein_ultracold_2007, gross_quantum_2017, krinner_two-terminal_2017, brantut_thermoelectric_2013, krinner_observation_2015, gruss_energy-resolved_2018}, trapped ion~\cite{gardiner_quantum_2014, blatt_quantum_2012, schneider_experimental_2012}, and quantum dot and other platforms~\cite{hanson_spins_2007, krantz_quantum_2019, hanson_room-temperature_2006}. This will, in turn, provide new insights into, e.g., quantum effects in thermodynamic  machines~\cite{deffner_quantum_2019} (i.e., quantum engines~\cite{deffner_nonequilibrium_2011, abah_single-ion_2012}) and may lead to scalable, physical simulators. 

Macroscopic reservoirs are often taken as continuum, non--interacting collections of electrons or phonons. These {\em Gaussian environments} are exactly solvable when not in contact with the system. However, the combination of the reservoirs and the quantum system will typically lead to analytically intractable equations of motion. Numerical techniques are thus the method of choice to benchmark approximations or to find the solution---e.g., the non--equilibrium steady state (NESS) of the system or the real--time dynamics of the current, among other quantities. Even within this simplified setting, though, simulation is difficult since it must capture both the many--body nature of the system and the non--Markovian behavior due to contact with the environment. 

In order to address this challenge, there is a rapidly growing body of methods that exploit the Gaussian nature of the environment to move the system--environment boundary. Instead of a clean split between system and environment, the boundary is between the system and a finite---and, due to computational constraints, relatively modest---set of reservoir modes on one side and the remaining external environment---the ``true'' reservoir---on the other. We dub these {\em extended reservoir approaches} (ERAs) as the finite collection of modes acts as an extension of the reservoir into the system. These modes impart explicit non--Markovianity to the dynamics. This idea builds on the concept of pseudo-modes~\cite{imamoglu_stochastic_1994,garraway_decay_1997,garraway_nonperturbative_1997,zwolak_dynamics_2008}, where external, time-local (Markovian) damping broadens the modes into Lorentzian peaks turning a discrete reservoir into an effective continuum.

ERAs can be applied with fermionic (e.g., electronic reservoirs) or bosonic (e.g., thermal baths) environments, as well as with a variety of relaxation methods (inter-- and intra--mode relaxation, Markovian or non--Markovian relaxation, etc.). Almost all of the approaches to date have focused on the continuous relaxation (CR) regime, where relaxation is always present in the equation of motion. This framework can handle both non--interacting and  many--body systems, where the formal transport solution is given by Meir--Wingreen--like equations~\cite{gruss_landauers_2016,elenewski_communication_2017,gruss_communication_2017,zwolak_comment_2020,zwolak_analytic_2020}. However, due to practical limitations, most simulations are of non--interacting  electron~\cite{sanchez_molecular_2006,subotnik_nonequilibrium_2009,dzhioev_super-fermion_2011,ajisaka_nonequlibrium_2012,ajisaka_nonequilibrium_2013,zelovich_state_2014, zelovich_moleculelead_2015, schwarz_lindblad-driven_2016, zelovich_driven_2016, hod_driven_2016, zelovich_parameter-free_2017, chiang_quantum_2020} (summarized in  Ref.~\cite{elenewski_communication_2017}) or classical thermal transport~\cite{velizhanin_driving_2011, chien_tunable_2013,velizhanin_crossover_2015,chien_thermal_2017, chien_topological_2018}. Tensor networks, though, are now enabling the implementation of this framework to non--perturbatively simulate many--body impurities~\cite{arrigoni_nonequilibrium_2013, dorda_auxiliary_2014, dorda_auxiliary_2015, dorda_optimized_2017, chen_Markovian_2019,fugger_nonequilibrium_2020,lotem_renormalized_2020,wojtowicz_open-system_2020} and extended systems~\cite{brenes_tensor-network_2020}.

Recently, an idea was put forward to periodically refresh the reservoirs rather than provide continuous relaxation~\cite{purkayastha_periodically_2021}. This periodic refresh (PR) approach evolves the system and the finite collection of reservoir modes coherently except at periodic intervals where the reservoir modes are reset into their initial (isolated) equilibrium. The refresh time has a firm physical basis: The Fermi velocity dictates how fast the density wavefront reaches the finite boundary of the explicit reservoir~\cite{chien_landauer_2014}, which is more generally related to a Lieb--Robinson bound~\cite{purkayastha_periodically_2021,lieb_finite_1972}. PR is recent and, as such, its accuracy as a simulation technique has not yet been carefully investigated.

Within the CR approach, the accuracy is tied to the reproduction of the continuum spectral function---particularly the weighted spectral function for quantum transport. This also provides bounds on parameters. For instance, the relaxation strength has to be less than the thermal relaxation~\cite{gruss_landauers_2016,elenewski_communication_2017}, which yields a minimum reservoir size as well. Ultimately, though, convergence depends on details of the model. on one hand, anomalous behavior (at zero bias~\cite{gruss_landauers_2016} or with interference~\cite{chiang_quantum_2020} or weak coupling~\cite{wojtowicz_dual_2021}) can hinder convergence and also modify optimal parameter selection~\cite{elenewski_performance_2021}. On the other hand, saturation and smooth behavior of observables like current can enhance convergence beyond what hard bounds would suggest~\cite{wojtowicz_open-system_2020,wojtowicz_dual_2021}. A comparison of different ERAs is thus imperative for both understanding the machinery and determining the optimal approach.

\begin{figure}[t!]
    \includegraphics[width=\columnwidth]{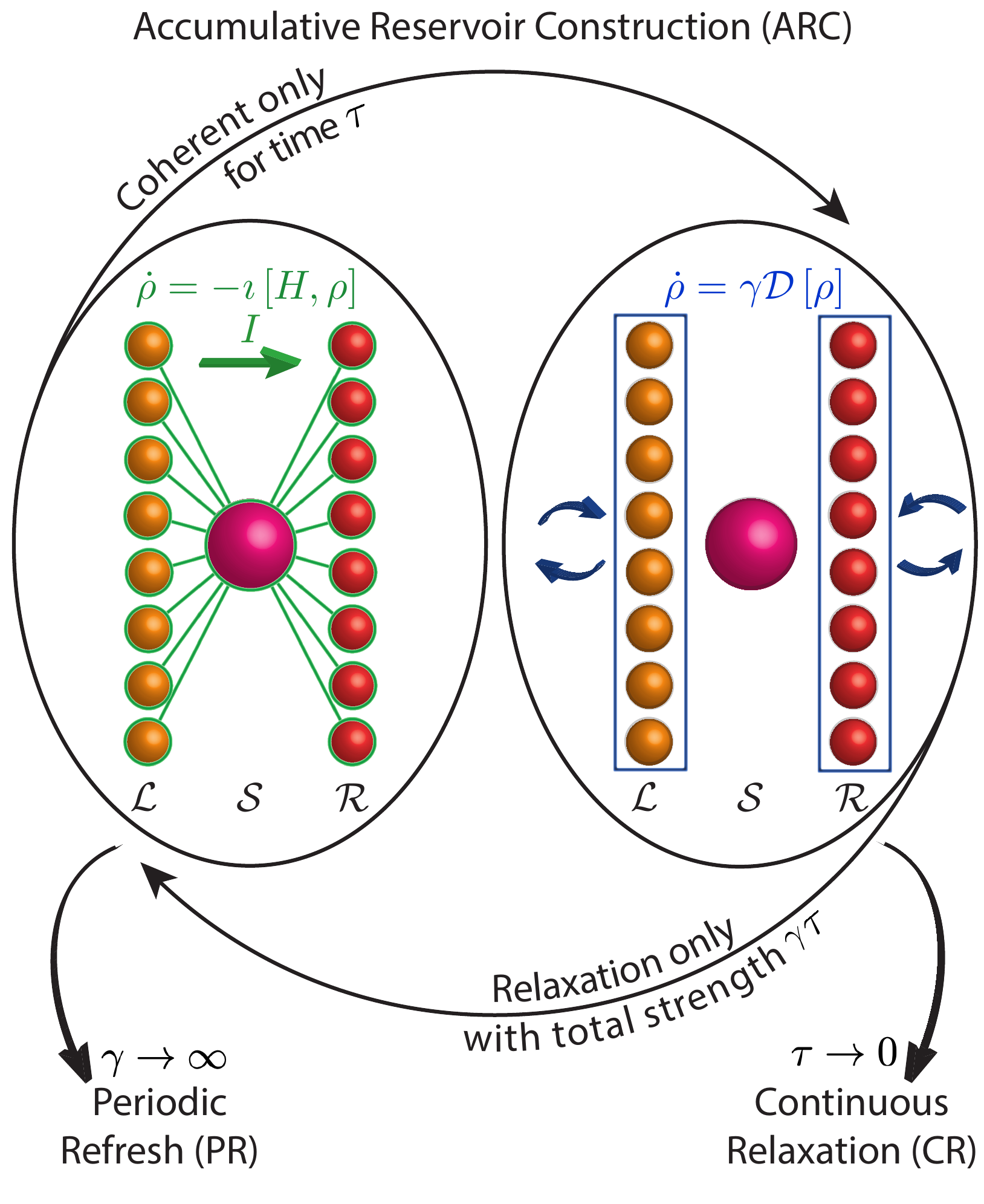}
    \caption{{\bf Accumulative reservoir construction (ARC).} ARC bridges two different approaches to simulate open quantum systems, from continuous relaxation (CR) to periodic refresh (PR). It consists of two steps of time $\tau$. The first is a unitary evolution with the composite Hamiltonian (highlighted in green on the left) of the system $\qs$ and reservoirs $\ql\qr$, the reservoirs being discretized into a finite number of modes $\NW$. The second step is a dissipative process that partially relaxes the reservoir modes towards their isolated equilibrium states (highlighted in blue on the right). In our implementation the dissipation is Lindblad damping with a homogeneous, intramode relaxation strength $\gamma$ and acts only on the reservoirs. ARC is shown in the context of quantum transport, where two reservoirs at different chemical potentials and/or  temperatures drive a current $I$ through an impurity or extended system $\qs$. Depending on $\gamma$, $\tau$, and $\NW$, the NESS from ARC can be a good approximation to the NESS from macroscopic reservoirs. CR corresponds to the limit $\tau \to 0$ at fixed $\gamma$, whereas PR corresponds to $\gamma \to \infty$ at fixed $\tau$.}
    \label{fig:1}
\end{figure}

Here, we take the first step in comparing distinct ERA approaches by providing a link between CR and PR approaches through what we call the {\it accumulative reservoir construction} (ARC), see Fig.~\ref{fig:1}. ARC reduces to CR and PR in two different limits. The core idea of the ARC is to divide the evolution into two steps: a unitary evolution of the system in contact with finite reservoirs followed by a {\em partial} relaxation of the reservoir modes (as opposed to full refresh in PR). We perform an extensive numerical characterization of ARC, focusing on the NESS of a solvable example, implemented both exactly and with tensor networks. A phase diagram results, where a ``cut'' separates CR-- and PR--like behavior. 
The point for the best estimate of the current jumps from the CR-- to the PR--like side of the cut, creating discontinuities in the operator entanglement and other quantities.
On the CR side, there is a modest level of entropy (i.e., low computational cost) but a moderate level of error (in current or system density matrix). On the PR side, there is a much larger entropy but much smaller errors. These regimes have different scaling behavior and thus the method of choice  depends on the computational resources available. This is manifest within tensor networks, where the PR--like regime of ARC is difficult to converge.

\section{Overview}

{We formulate the accumulative reservoir approach (ARC) as a method to numerically investigate quantum transport (Sec.~\ref{sec:framework}), as well as non--Markovian open quantum systems more generally. The ARC bridges two existing approaches using continuous relaxation (CR), and periodic refresh (PR). As this bridge, it has three core parameters: reservoir size, relaxation strength, and time between refresh events. We focus on the non--interacting resonant level model (Sec.~\ref{sec:model}) and extensively characterize the behavior of transport in this parameter space---i.e., we provide a phase diagram---for this model (Sec.~\ref{sec:results_arc}). This allows us to benchmark ARC against exact results to assess the convergence of the steady--state current, as well as the impurity state. We then conclude (Sec.~\ref{sec:conclusions}).}

{The phase diagram for ARC has two overarching categorizations. The first is whether the behavior is CR-- or PR--like. There is a ``cut'' that defines the transition between the two and signifies the interplay of ARC parameters with different physical timescales. The second categorization is Kramers' turnover. ARC broadly falls into the paradigm of this turnover. The current has over-- and under--damped regions of behavior. The overdamped behavior is Zeno--like regardless of whether one is in the CR-- or PR--like regime, as it is essentially constant measurement from the relaxation that dictates the current. Underdamped behavior manifests itself differently in the CR-- and PR--like regimes. Except at the cut, the region between over-- and under--damped behavior captures the physics of the current. This region grows as the number of reservoir modes increases and is thus the physically applicable regime that reproduces the continuum limit.}

{In the context of efficient simulation, CR and PR yield different error--cost scaling, with PR having a faster decay of error versus cost but it also has a larger prefactor. The larger prefactor causes convergence difficulties. With the tensor network implementation we employ, CR is thus preferable. ARC, however, has a large swath of phase space between CR and PR. The intermediate regime is refresh--like in its physics, but has even better error--cost scaling. It still is more challenging to converge than CR, but may serve as a target for more accurate  simulations.}

{We expect the results and conclusions for ARC to generalize to more complex models, including those with many--body interactions. For CR, for instance, the overdamped regime can be derived directly from Meir--Wingreen--like equations for both non--interacting and many--body impurities~\cite{gruss_landauers_2016}. Those equations also show that CR limits to the exact result. The underdamped regime has been proven for a large class of models and is expected to always hold based on physical arguments. Similarly, physical arguments suggest the turnover for PR and PR--like behavior will always hold. By construction, PR also limits to the right current in the intermediate turnover regime. Many--body interactions may change the error--cost scaling, especially its non--asymptotic behavior (i.e., smaller explicit reservoir sizes) by introducing different time and temperature scales, or power--law decaying correlations. We do not expect these will change the qualitative distinctions in the error--cost relationships between CR, PR, and intermediate regimes. These distinctions are due to balancing the reservoir representation and relaxation--induced reduction of entanglement, neither of which are system dependent.}

\section{Framework}
\label{sec:framework}

While the ERA is a general approach to handle a large class of non--Markovian open quantum systems, we will employ it here in the context of quantum transport through an impurity or junction region. In this scenario, a chemical potential (or temperature) drop between two (non--interacting) metallic electronic reservoirs $\ql$ and $\qr$ will drive a current through a system $\qs$, as seen by the left schematic in Fig.~\ref{fig:1}. The total Hamiltonian, 
\begin{equation}\label{eq:hlsr}
     H =  H_{\qs} + H_{\ql} + H_{\qr} + H_{\qi} ,
\end{equation}
is given by a sum of a system Hamiltonian $H_{\qs}$, $\alpha\in\{\ql,\qr\}$ reservoir Hamiltonian, 
\begin{equation}\label{eq:hlr}
 H_{\alpha} = \sum_{k \in \alpha} \omega_{k}\cid{k} \ci{k},
\end{equation}
and coupling Hamiltonian between $\qs$ and $\ql\qr$, 
\begin{equation}\label{eq:hinteraction}
     H_{\qi} =  \sum_{k\in \ql\qr }  \sum_{i \in \qs}  \left( v_{ki} \cid{k} \ci{i} + v_{ik} \cid{i} \ci{k} \right) .
\end{equation}
Here, $\cid{m}$ ($\ci{m}$) is the fermionic creation (annihilation) operator for mode $m \in \ql \qs \qr$, the index $m$ carries all necessary labels (frequency, location, spin, etc.), and all Hamiltonians are in terms of frequencies. Each reservoir is taken to have $N_\ql = N_\qr = \NW$ modes, where the label $\qw$ signifies the electronic band (the two reservoirs could have a different number of modes but we do not consider that case). The Hermitian coupling matrix $v_{ki}=v_{ik}^\ast$ defines the quadratic interaction between the reservoirs and the system. More generally, $H_{\qi}$ could be taken as a coupling linear in the reservoir operators but with an arbitrary system operator. We restrict to the transport setting where there is just hopping between the reservoirs and system. The Hamiltonian $H_\qs$ is also generally arbitrary, potentially having, e.g., electron--electron or electron--vibrational couplings. In our example, as described later, we take it to be quadratic so that we can compare to the exact solution.
 
In addition to the Hamiltonian, we must have proper boundary conditions to drive a current. There are many ways to do this, such as having an initially disconnected reservoirs $\ql$ and $\qr$ with a chemical potential drop between them or by turning on a chemical potential drop at an initial time where $\ql\qs\qr$ is in equilibrium state, among others. For continuum (macroscopic) reservoirs, as well as ERA, we will take the former approach. The state at time $t=0$ is thus
\begin{align}
\rho(t=0)=\rho_\qs(0) \otimes \rhoI{\ql} \otimes \rhoI{\qr},
\end{align}
where we take $\rho_\qs(0)$ to be the maximally mixed state on $\qs$ and $\rhoI{\alpha}$ the isolated equilibrium state of reservoir $\alpha$. The NESS can, in principle, depend on initial conditions but the example we will examine in this work has a unique steady state. The states $\rhoI{\alpha}$ are defined by the Fermi--Dirac distribution,
\begin{align}
f^{\alpha}(\omega)=\frac{1}{1+e^{\beta_\alpha(\omega-\mu_\alpha)}},
\end{align}
at chemical potential $\mu_\alpha$ and thermal relaxation time $\beta_\alpha = \hbar/k_B T_\alpha$, where $\hbar$ is the reduced Planck's constant, $k_B$ is Boltzmann's constant, and $T_\alpha$ is the temperature. {Note that, for convenience, we work with $\beta$ as the inverse thermal relaxation rate, i.e., thermal relaxation time, which has a unit of time.} We will take a symmetrically applied bias $\mu_\ql = -\mu_\qr = \mu / 2$ and uniform temperature $T_\ql=T_\qr=T$. The ERA relaxes the reservoirs toward $\rhoI{\alpha}$ in order to maintain a chemical potential and/or temperature imbalance between the finite collection of explicit reservoir modes.

For continuum reservoirs, $\NW \to \infty$ and $v_{ki}\sim1/\sqrt{\NW}$, the NESS is   
\begin{align}
\rhoo = \lim_{t\to\infty} \Big(\lim_{\NW\to\infty} \tr_{\ql\qr}\left[ e^{-\im t  H } {\rho}(0)  e^{\im t  H}\right]\Big) ,
\end{align}
where $\tr_{\ql\qr}$ denotes the trace over $\ql\qr$. A hallmark of a NESS is that there are currents flowing due to the different chemical potentials or temperatures. The goal of ERA is to supply an efficient numerical technique to find a NESS that has a well--defined convergence to $\rhoo$, while simultaneously enabling the numerically exact inclusion of many--body interactions. This enables determining the current--voltage relationship or energy currents, as well as other characteristics of the system. 

When working with non--interacting, number-conserving reservoirs, Eq.~\eqref{eq:hlr}, linearly coupled to the system, as in Eq.~\eqref{eq:hinteraction}, the NESS is governed by the reservoir spectral functions. Commonly, the reservoirs are coupled each to only a single mode in the system, $v_{k i} \equiv v_k$ (the system mode $k$ can be different for different $\alpha$). In this case, the spectral function is 
\begin{equation}\label{eq:Jw}
J_\alpha(\omega)=2\pi \sum_{k \in \alpha} v_{k}^2 \delta(\omega-\omega_k) . 
\end{equation}
More generally, the spectral function will be a matrix that characterizes the couplings to all $\qs$ modes~\cite{gruss_landauers_2016,zwolak_analytic_2020}. For the specific model we examine, each reservoir is only connected to a single boundary site on the left (for $\ql$) and right (for $\qr$) of the system. Thus, Eq.~\eqref{eq:Jw} will be the relevant quantity. All microscopic models of the reservoirs that lead to the same macroscopic ($\NW \to \infty$) spectral functions give the same dynamics and NESS. This feature is often used to obtain finite-size reservoirs that systematically accumulate essential features of these functions when increasing the number of reservoir modes~\cite{zwolak_finite_2008,chin_exact_2010,binder_reaction_2018,elenewski_performance_2021}. The approaches we discuss are all based on such finite-size representations of the reservoirs. 

Since ERA methods use a finite number of modes that are in turn open to implicit environments, we need to specify how these implicit environments relax the explicit modes. In addition to basic model properties (e.g., the spectral function and $\qs$), this relaxation will dictate the NESS and how well the technique approximates $\rhoo$. We will now describe the two limits of ARC, one with damping given by continuous, Markovian relaxation and the other with periodic, full refreshes. We then will turn to the general ARC description.
\subsection{Continuous relaxation (CR)}
\label{sec:cr}

For CR, each reservoir is damped continuously via a Lindblad dissipator. The equation of motion is 
\begin{equation} \label{eq:evolution_cr}
\dot{\rho} = - \im [H, \rho] + \gamma \qd [ \rho ] ,
\end{equation}
where the dissipative contribution is 
\begin{align} \label{eq:dissipator}
\qd [ \rho ] =
    & \sum_{k\in\ql\qr}  f^{\alpha_k}(\omega_k) \left( \cid{k} \rho \ci{k}
        - \frac{1}{2} \left \{ \ci{k} \cid{k}, \rho\right \}\right) \notag \\
    + & \sum_{k\in\ql\qr} \left[1- f^{\alpha_k}(\omega_k)\right] \left( \ci{k} \rho \cid{k}
        - \frac{1}{2} \left \{ \cid{k} \ci{k}, \rho \right \} \right)
\end{align}
and $[\cdot,\cdot]$ ($\{ \cdot,\cdot \}$) is the commutator (anticommutator). This form, Eq.~\eqref{eq:evolution_cr} and Eq.~\eqref{eq:dissipator}, assumes a homogeneous relaxation strength $\gamma$ (this assumption can be relaxed but it makes little difference with common discretizations of the reservoirs~\cite{elenewski_performance_2021}) and only intramode relaxation. We can alternatively cast the equations in terms injection and depletion rates, 
$\gamma_{k+}=\gamma f^{\alpha_k}(\omega_k)$ and $\gamma_{k-}=\gamma \left[1- f^{\alpha_k}(\omega_k)\right]$, respectively. Without the coupling to $\qs$, each mode thermalizes to its own equilibrium state set by the temperature $T_\alpha$ and chemical potential $\mu_\alpha$ of its respective reservoir $\alpha$. 

{We note that Eq.~\eqref{eq:dissipator} is Markovian for $\ql\qs\qr$ since the reservoirs $\ql\qr$ are coupled to Markovian environments. The dynamics for $\qs$ only, though, is non--Markovian. Each explicit mode in $\ql$ or $\qr$ that has relaxation will have a Lorentzian spectral function and thus have a memory time---i.e., non--Markovianity---of $1/\gamma$. From $\qs$'s perspective, multiple explicit modes in $\ql\qr$ add frequency dependence to the environment's response (i.e., ``color'') within this $1/\gamma$ timescale. Moreover, in order to target physical dynamics, one takes the joint limit of $\NW \to \infty$ and $\gamma \to 0$, resulting in the memory time going to infinity and capturing the full character of the continuum.}

While Eq.~\eqref{eq:dissipator} can be derived for weak coupling to external environments with chemical potentials at $\pm \infty$~\cite{elenewski_communication_2017}, the important aspect is that its NESS converges to the continuum, macroscopic result as $\NW \to \infty$ and then $\gamma \to 0$~\cite{gruss_landauers_2016}. The formal solution provides the guarantee of convergence for both non--interacting and many--body systems~\cite{gruss_landauers_2016,elenewski_communication_2017,gruss_communication_2017,zwolak_comment_2020,zwolak_analytic_2020}, as well as bounds on parameters (such as the aforementioned {$\beta \gamma \le 1$})~\cite{gruss_landauers_2016,elenewski_communication_2017}. Yet, there are still many aspects of the convergence, such as its rate and the presence of features, that need to be understood and properly handled~\cite{elenewski_performance_2021,wojtowicz_dual_2021}. The goal is to obtain accurate, physically meaningful results with minimal computational resources (often, the smallest $\NW$ possible but other characteristics are equally important). 

Toward this end, one can explore the NESS, and specifically the steady--state particle current, from Eq.~\eqref{eq:evolution_cr} treating $\gamma$ (and $\NW$) as parameters. Generally there are three basic regimes analogous to Kramers' turnover in solution--phase chemical reactions~\cite{kramers_brownian_1940} as the relaxation strength varies~\cite{gruss_landauers_2016} (similarly for the same ERA approach for classical thermal transport~\cite{velizhanin_crossover_2015}):
\begin{itemize}
    \item {\em Small--$\gamma$} or {\em relaxation--limited} regime. In this regime, the particle current is dictated by the relaxation strength. The supply of particles by the implicit environments is too small compared to the intrinsic current---the physical current---that can flow through the system. This regime can be cast as a three resistor model, with two resistors at the interfaces to the left and right external environments and one resistor for $\ql \qs \qr$. The latter is much smaller than the former and thus the interfaces to the external environments dominate (albeit the former is proportional to $1/\gamma \NW$ and thus the transition from relaxation--limited to the intrinsic current depends on $\NW$). Thus, there is a linear increase of the total current versus $\gamma$. This regime is also characterized by overly coherent interactions within $\ql \qs \qr$, i.e., that there are unphysical global oscillations within all the explicit degrees of freedom. {From the initial state, a wavefront transverses $\ql\qs\qr$ with the Fermi velocity in the same way as in a closed system. This transversal and associated oscillations decays into the steady state, but they still  influence the character of the NESS.}
    \item {\em Intermediate--$\gamma$} or {\em physical} regime. In this regime, the supply rate is sufficiently strong to support the intrinsic, physical current without disturbing it significantly. As with the small $\gamma$ regime, one can cast this regime within the three resistor model. But now, since $\gamma$ is larger, the $\ql \qs \qr$ resistor dominates and one obtains the physical current from calculations. In this regime, a plateau in the current develops and increases in size with $\NW$. It is in this regime that the current---typically the most important observable---accurately captures the current in the true NESS. The approximate NESS will also approach $\rhoo$~\cite{elenewski_performance_2021}. Extending this plateau is the practical realization of the limits $\NW \to \infty$ and then $\gamma \to 0$ (i.e., as $\NW$ increases but is still finite, adjusting $\gamma$ to be sufficiently small but not so small to turn over into the small $\gamma$ regime).
    \item {\em Large--$\gamma$} or {\em overdamped} regime. In this regime, the relaxation is too strong, suppressing the coherence between the extended reservoir and $\qs$ that is necessary to carry a current (as currents are just coherences). This effectively decouples the extended reservoirs, resulting in the current decaying as $1/\gamma$. The situation here is analogous to the Zeno effect. As a consequence of the decoupling, $\qs$ itself is overly coherent. As with the other two regimes, one can also cast this regime---in fact, all three regimes---into a series of resistors. One has to include two additional resistors, though, one at the interface between $\ql\qs$ and one at $\qs\qr$ (but with resistor proportional to $\gamma$ rather than $1/\gamma$). 
\end{itemize}

While the above regimes hold generally, there can also be additional regimes due to anomalous behavior~\cite{wojtowicz_dual_2021}. For example, at finite relaxation strength, the Markovian relaxation fails to reproduce the Fermi level, increasing the transmission across $\ql\qs\qr$ due to artificial spreading of the occupied density of states. This is related to the bound {$\beta \gamma \le 1$} and, as such, it manifests at strong relaxation ({\em a Markovian anomaly}). For proportionally coupled reservoirs (i.e., ones where we have the same set of modes in $\ql$ and $\qr$), another anomaly can occur at weak relaxation due to virtual (resonant) transitions ({\em a virtual anomaly}). Small intrinsic currents, e.g., for weak system--reservoir coupling, reveal these anomalies. The plateau forms between them as $N_\qw$ increases. We previously proposed a heuristic approach to estimate the optimal $\gamma$ to target the physical regime for a general, including many--body, $\qs$~\cite{elenewski_performance_2021}. The approach removes the virtual anomaly by applying a small shift between the frequencies of $\ql$ and $\qr$ modes. The estimate for $\gamma$ is where the turnover curve intersects the curve without this shift, i.e., it finds a domain of confidence for CR (a sweet spot for $\gamma$). We will employ this approach in this work for CR. Additional information on this procedure is in App.~\ref{si:cr}.

\subsection{Periodic refresh (PR)}\label{sec:pr}

In the PR approach~\cite{purkayastha_periodically_2021}, the system and finite-size reservoirs evolve fully unitarily for a time $\tau$ after which the reservoirs are ``refreshed'' (reset to their initial, isolated equilibrium state), and the process repeats periodically. At cycle $p=0,1,\ldots$, the evolution follows 
\begin{equation}\label{eq:evolution_pr}
    {\rho}(p\tau+\tau)  = \tr_{\ql\qr}[e^{-\im \tau  H } \rho(p \tau)  e^{\im \tau  H}] \otimes \rhoI{\ql} \otimes \rhoI{\qr} .
\end{equation}
This provides the state of $\ql \qs \qr$ at discrete intervals $p \tau$. In order to examine the behavior of this approach, we will also sometimes make use of the state in between these discrete intervals at times $t=p \tau + \tau^\p$ with $\tau^\p < \tau$. When we do, we employ the assignment, $\rho(p \tau + \tau^\p) = e^{-\im \tau^\p  H } \rho(p \tau)  e^{\im \tau^\p  H}$ (in words, only coherent evolution is included in assigning the time).  A physical realization of this process would either require multiple copies of each reservoir---a so--called collisional or repeated interaction model---or the ability to otherwise control and reset the reservoir (e.g., turn off interactions with and in the system and ``refill'' the explicit reservoir modes). 

The NESS in PR is a periodic state.  However, if we stroboscopically look at the system at time steps of $\tau$, $\rho_\qs (p \tau) = \tr_{\ql \qr}[ \rho(p \tau)]$, then we will find an approximation to $\rhoo$ as $p\to\infty$. The interfacial currents (between the reservoirs and system) are often of interest, e.g., for special cases, such as impurities with only one site in contact with the reservoirs, they are the only place to obtain the overall current. In which case, one uses $\rho(p \tau - \epsilon)$ with $\epsilon \to 0^+$. This selects the full state just before the refresh. It has no bearing on the estimated $\rhoo$, as $\rho_\qs (p \tau) = \lim_{\epsilon \to 0^+} \tr_{\ql\qr}[\rho(p \tau - \epsilon)]$ since the refresh does nothing to the $\qs$'s state. It does, however, retain the system--reservoir correlations, enabling an estimate of the steady--state interfacial current. 

In a similar manner to CR, as $\NW \to \infty$ and then $\tau \to \infty$, the system's NESS will converge to the continuum, macroscopic NESS. The periodic refreshing, though, stabilizes the state into a true NESS, opposed to  closed---``microcanonical''---evolution (where the system state and/or current is taken from the quasi-steady state that forms at finite time~\cite{ventra_transport_2004,bushong_approach_2005,sai_microscopic_2007,chien_bosonic_2012,chien_interaction-induced_2013,chien_landauer_2014,gruss_energy-resolved_2018}). 
We further note that, for a given $\NW$, a physically meaningful $\tau$ results from a lattice mapping of the reservoir spectral function followed by Lieb--Robinson arguments. For, e.g., a simple one--dimensional (1D) reservoir, this $\tau$ is
\[ \label{eq:PhysicalTau}
\TW = \NW / 2\omega_0 ,
\]
where $2 \omega_0$ is the Fermi velocity, which is set by the (uniform) hopping frequency, $\omega_0$, in the 1D lattice. This $\tau$ signifies when the current wavefront has impinges of the boundary of the finite reservoir. This effectively yields only one free parameter (either $\NW$ or $\tau$, both increasing in tandem). Just as with CR the non--Markovian character increases as this free parameter increases. We designate $\TW$ the physical timescale. Additional information on PR, specifically its Kramers' turnover--like physics that we study in this work, is in App.~\ref{si:pr}.

\subsection{Accumulative reservoir construction (ARC)}\label{sec:ARC}

As with PR, the ARC also takes a periodic process starting first with a fully unitary evolution of the system and explicit reservoir modes for time $\tau$. This is followed by a dissipative process of total strength $\gamma \tau$ with the system and system--reservoir coupling frozen. Since these processes are alternating, the relaxation can be applied for time $\tau$ with strength $\gamma$ or could be an ``impulse'' or discrete ``gate'' with the same total strength $\gamma \tau$.  This cyclical process is depicted in Fig.~\ref{fig:1}. At cycle $p=0,1,\ldots$, the evolution follows the form 
\begin{equation}\label{eq:evolution_arc}
    {\rho}(p\tau+\tau)  = e^{\gamma \tau \qd}[ e^{-\im \tau  H } \rho(p \tau)  e^{\im \tau  H}] ,
\end{equation}
where $e^{\gamma \tau \qd} [\,\cdot\,]$ is the evolution superoperator for the dissipation from Eq.~\eqref{eq:dissipator}. Note that there are two separate evolutions in Eq.~\eqref{eq:evolution_arc} each for a time $\tau$, yet we only increment the time of the density matrix by $\tau$ (opposed to $2 \tau$). In other words, we do not count the dissipative evolution in the time. As with PR, we employ the assignment, $\rho(p \tau + \tau^\p) = e^{-\im \tau^\p  H } \rho(p \tau)  e^{\im \tau^\p  H}$, for intermediate times $t=p \tau + \tau^\p$ with $\tau^\p < \tau$. 

For ARC, the coherent evolution is exactly as for PR. The dissipation, though, does not give a complete refresh of the reservoirs. Rather, it relaxes the reservoirs with a total strength of $\gamma \tau$. If $\tau$ is fixed, for instance, but $\gamma \to \infty$, then this performs a full refresh and thus completely recovers PR. If one fixes $\gamma$ but takes $\tau \to 0$, then one recovers CR. This limit is just a first order Trotter expansion of the evolution in Eq.~\eqref{eq:evolution_cr}. This motivates leaving out the dissipative evolution in the assignment of intermediate times, as the two processes taken separately with a step $\tau$ is a numerical approximation to a single step $\tau$ taken jointly. As we will see, even away from the CR or PR limits, there is a wide range of parameters where the NESS approximates $\rhoo$. 

This discussion already makes apparent that there are three free parameters in ARC ($\gamma$, $\tau$, and $\NW$). This is opposed to CR and PR, which both have an effective single parameter. Albeit, in the case of CR, imposing a single relation between $\gamma$ and $\NW$ will not be the most effective use of computational resources; Rather, one wants to make use of the existence of a domain of confidence to set the relationship between the two~\cite{elenewski_performance_2021}{, yielding $\gamma$ for a set $\NW$}. A similar consideration holds for PR. One can use the physical $\TW$, but this may not be the most effective in terms of error for a given cost, as we will discuss. For ARC, however, using $\gamma$ and $\tau$ as the parameters is not always the most suitable choice. Rather, the product, 
\begin{equation}
\gamma \tau = \mathrm{Total \, relaxation},
\end{equation}
provides the total amount of relaxation at each cycle and is the only relevant quantity for the dissipation. For the other parameter, we choose the quantity
\begin{equation}\label{eq:action}
    \action = \bigg|\tau-\im \frac{2}{\gamma}\bigg| = \mathrm{Action},
\end{equation}
which is the magnitude of a complex action. This choice is motivated by the way in which the relaxation appears in the retarded Green's function, see Ref.~\cite{gruss_landauers_2016}. We will numerically characterize the ARC as function of these parameters using a solvable example. We will see that the choice of parameters makes many features of the phase diagram for ARC readily apparent, as well as has the two limits (CR and PR) as ``Cartesian'' boundaries on polar opposites of the phase space. These aspects make it a good choice, but there may be other choices that are superior. For instance, the bare parameters, $\tau$ and $\gamma$, remain meaningful when interpreting aspects of the evolution, as we will discuss. 

Finally, we can view ARC (and PR) as an example of a bang-bang protocol~\cite{cerezo_variational_2021, yang_optimizing_2017}.  Such protocols appear, for instance, in the context of quantum computation as a procedure outputting the desired state through the alternating application of two (or more) non--commuting processes. In variational quantum algorithms~\cite{cerezo_variational_2021, yang_optimizing_2017}, protocol parameters are optimized, typically to minimize some cost function, such as energy. Similarly, in ARC, the parameters should, within finite resources, best reproduce the influence of continuum reservoirs. As the latter does not admit a straightforward variational form, we approximate the optimal ARC parameters based on a physical understanding of the process. Appendix~\ref{si:arc_kramers} contains additional discussion on ARC.

\begin{figure}[t!]
\includegraphics[width=\columnwidth]{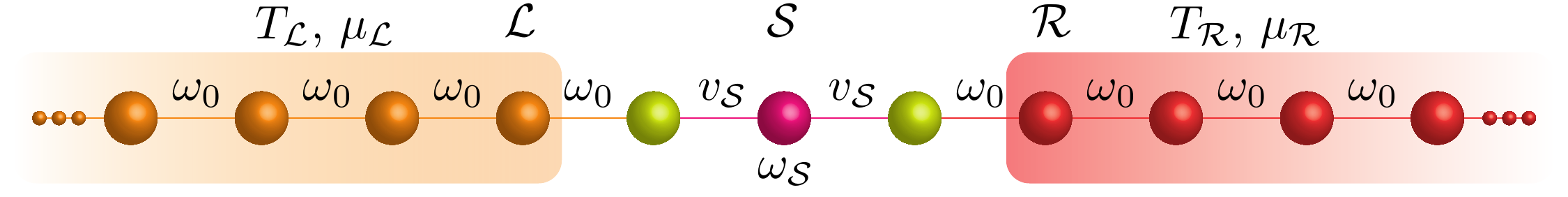}
\caption{{\bf Resonant--level model.} The schematic shows the example---the resonant level model---used to numerically characterize ARC. We will use the parameters $\onS = v_\qs = \omega_0/2$, several different temperatures but with $T_\ql=T_\qr=T$, and the chemical potentials $\mu_\ql = -\mu_\qr = \omega_0/4$ (i.e., a total potential drop of $\omega_0/2$ applied symmetrically). By going to the single particle eigenbasis of $\ql$ and $\qr$, the model conforms to the setup in Fig.~\ref{fig:1}. Even within tensor networks, we simulate the reservoirs using their eigenbasis and thus their spatial dimension is not directly relevant.}
\label{fig:2}
\end{figure}

\section{Example system}
\label{sec:model}

In all the results below, we consider reservoirs that are 1D lattices with a uniform nearest--neighbor hopping frequency $\omega_0$. The impurity region is composed of $\NS=3$ sites also arranged in a one--dimensional configuration with Hamiltonian,
\begin{equation}\label{eq:RLM}
     H_\qs = \onS  a_2^\dagger  a_2 + v_\qs ( a^\dagger_1 a_2 +  a^\dagger_2 a_3 + \mathrm{h.c.}), 
\end{equation}
where $ a_i$ ($ a^\dagger_i$) are fermionic annihilation/creation operators for site $i$ in $\qs$ and h.c.~signifies the Hermitian conjugate terms. As the example model, we take the onsite frequency of the middle site to be $\onS=\omega_0/2$ and hopping amplitude $v_\qs= \omega_0/2$. As indicated in Fig.~\ref{fig:2}, the reservoirs $\ql$ and $\qr$ will both be at the initial temperature $T$, while the (symmetric) chemical potential imbalance of $\mu_\ql = -\mu_\qr = \omega_0/4$ drives the particle current. This model is a non--interacting resonant level model (see Refs.~\cite{boulat_twofold_2008,bidzhiev_out--equilibrium_2017} for the interacting resonant level model). For the reservoirs, by choosing a finite lattice length of $\NW$ for $\ql$ and $\qr$ and then diagonalizing the single--particle Hamiltonians, the reservoirs and the system--reservoir couplings are cast in the form of Eq.~\eqref{eq:hlr} and Eq.~\eqref{eq:hinteraction}. This process provides the finite reservoirs that we use. For $\NW \to \infty$, the reservoir spectral function for a uniform 1D lattice with hopping $\omega_0$ (in contact with the impurity boundary with coupling also $\omega_0$) are
\begin{equation} \label{eq:ourspectralfun}
J_\ql(\omega)=J_\qr(\omega)=J(\omega)=\sqrt{4 \omega_0^2 - \omega^2} .
\end{equation}
We are after the NESS of this model. Although there are no many--body interactions, we expect the qualitative nature of the results to carry over to presence of many--body interactions in the system $\qs$.

\subsection{Reference NESS}
\label{sec:RefResults}

Since this is a non--interacting model, we can solve for all properties of the system in the continuum, macroscopic limit by employing non--equilibrium Green's functions. It is helpful to write the Hamiltonian, Eq.~\eqref{eq:hlsr}, in terms of its single--particle counterpart, $\bH$,
\begin{equation}
H = \sum_{m,n\in\ql\qs\qr} \bH_{mn} \cid{m} \ci{n} ,
\end{equation}
as well as the system's Hamiltonian itself,
\begin{equation}
H_\qs = \sum_{i,j\in\qs} [\bHS]_{ij} \cid{i} \ci{j} ,
\end{equation}
where $[\bHS]_{ij} = \bH_{ij}$ is a submatrix. Using this, the retarded (advanced) Green's function of the system is
\begin{equation}
\bG^{r (a)} = 1/(\omega - \bHS - \bS^{r (a)}_\ql - \bS^{r (a)}_\qr) , 
\end{equation}
with respective self--energies
\begin{equation}
\bS^{r (a)}_\alpha=\sum_{k\in\alpha} g^{r (a)}_k \ok.
\end{equation}
We use $\kk$ for the coupling vector between mode $k\in\ql\qr$ and all sites $i\in\qs$, i.e., $\inner{i}{v_k}=v_{ik}$, and  $g_{k}^{r(a)}=1/(\omega-\omega_k \pm \im \eta)$ for the retarded (advanced) Green's function of $k\in\ql\qr$. As usual, the limit of $\eta \to 0^+$ is taken at the end of the calculation. Note that this formalism also permits the formal CR solution, including analytic in some cases, of the current and correlation matrices for $\ql \qs \qr$~\cite{gruss_landauers_2016,elenewski_communication_2017,gruss_communication_2017,zwolak_comment_2020,zwolak_analytic_2020}.

These quantities give the spectral densities (which are the spectral functions, the terminology depends on the community) $\bGa^{\alpha} = \im ( \bS^r_{\alpha} - \bS^a_{\alpha}) = -2 \Im \bS^r_{\alpha}$, 
\begin{equation} \label{eq:unweightedsigma_gen}
\bGa^{\alpha}(\omega) = \im\sum_{k\in \alpha} 
                       \left[g_{k}^{r}(\omega) - g_{k}^{a}(\omega) \right] \ok.
\end{equation}
The population--weighted counterpart is
\begin{equation}  \label{eq:weightedsigma_gen}
\tilde{\bGa}^{\alpha}(\omega) = \im\sum_{k\in \alpha} f^\alpha (\omega)
                       \left[g_{k}^{r}(\omega) - g_{k}^{a}(\omega) \right] \ok.
\end{equation}
With this, the current in the continuum, macroscopic limit is given by the Landauer expression, 
\begin{equation} \label{eq:nonintCurrStandard}
\Io = \int\frac{d\omega}{2\pi} \left( f^\ql (\omega) - f^\qr (\omega) \right) \tr \left[ \bGa^\ql \bG^r \bGa^\qr \bG^a \right] .
\end{equation}
In other words, this is the current in the true NESS, the one we are after. We designate it with the superscript $\circ$ to indicate that it is our standard, or reference, current. 

Since we are considering an example with a quadratic Hamiltonian, the NESS is Gaussian state and thus can be completely characterized by the correlation matrix 
\[
[ \cm^{\circ}_\qs ]_{ij} = \tr [ \cid{j} \ci{i} \rhoo ] ,
\] 
where $\cm^\circ_\qs$ is the system's submatrix of the full correlation matrix. 
The correlation matrix is often referred to as the single--particle density matrix. It is a Hermitian, semi-positive definite matrix, although it does not have a trace equal to one. We can similarly write the system's correlation matrix in the NESS, 
\begin{equation} \label{eq:nonintCorrStandard}
\cm^\circ_\qs = \int\frac{d\omega}{2\pi} \bG^r \left( \tilde{\bGa}^{\ql} + \tilde{\bGa}^{\qr}\right) \bG^a .
\end{equation}
Unlike the current, only the population--weighted spectral densities appear here. They are the source of particles that propagate with $\bG^r [\,\cdot\,] \bG^a$ into the system. 

While the expressions above, Eq.~\eqref{eq:nonintCurrStandard} and Eq.~\eqref{eq:nonintCorrStandard}, for the current and correlation matrix are more general, our example only has non--zero elements of the self--energies and the spectral densities for $i=1,j=1$ and $i=\NS, j=\NS$. For the self--energies, these are
\begin{equation}
[\bS^r_{\ql}(\omega)]_{11}=[\bS^r_{\qr}(\omega)]_{\NS \NS}=-\frac{1}{2}\left( \omega + \im \sqrt{4 \omega_0^2 - \omega^2}   \right) .
\end{equation}
The spectral densities (functions) readily follow, showing consistency with these quantities and Eq.~\eqref{eq:ourspectralfun}. 

\subsection{Non--interacting ARC}
\label{sec:NonintARC}
The evolution generated by Eq.~\eqref{eq:evolution_cr} with the quadratic Hamiltonian, Eq.~\eqref{eq:hlsr}, retains the Gaussianity of the state.
As a result, the evolution (which is also particle conserving) can be formulated via the correlation matrix,
\[
\cm_{mn} = \tr [ \cid{n} \ci{m} \rho ] , 
\]
with $m,n\in\ql\qs\qr$. In this form, we have~\cite{elenewski_communication_2017}
\begin{equation}
\dot{\cm} = -\im \left[\bH, \cm \right] + \gamma \qd [ \cm ],
\end{equation}
with the dissipation in Eq.~\eqref{eq:dissipator} translating to
\begin{equation}
\label{eq:CMdissipator}
\qd [ \cm ]_{mn} = f^{\alpha_m}(\omega_m) \delta_{mn} \delta_{m \in \ql \qr} - \cm_{mn} \frac{\delta_{m \in \ql \qr} + \delta_{n \in \ql \qr}}{2}.
\end{equation}
Here, the $\delta_{mn}$ is the normal Kronecker delta, and $\delta_{m \in \ql \qr}$ is 1 for the mode in one of the reservoirs and 0 otherwise, i.e., no relaxation occurs within the system.

Consequently, for purely unitary evolution over a time $\tau$ (without dissipation), the evolution is
\[
    e^{- \im \tau \bH} \cm  e^{ \im \tau \bH}.
\]
For pure dissipation for a time $\tau$ (without the action of the Hamiltonian), integrating Eq.~\eqref{eq:CMdissipator} leads to an update
\[
    \Gop \cm \Gop^\dagger + \Pop,
\]
where the dissipative matrices are
\[
\Gop_{mn} = \delta_{mn}  e^{- \gamma \tau \delta_{m \in \ql \qr} /2}
\]
and
\[
\Pop_{mn} =  \delta_{mn} \left( 1-e^{-\gamma\tau \delta_{m \in \ql \qr} } \right) f^{\alpha_m} (\omega_m) .
\]
For ARC in Eq.~\eqref{eq:evolution_arc}, the evolution is a unitary followed by a dissipation step. The cycle $p$ of this process is
\begin{equation} \label{eq:ARC}
    \cm(p\tau+\tau) = \mathbf{M} \cm(p \tau) \mathbf{M}^\dagger + \Pop,
\end{equation}
with
\[
\mathbf{M} = \Gop e^{- \im \tau \bH}.
\]
These results hold for any Gaussian system, i.e, in absence of many--body interactions. The system can be a lattice of arbitrary dimension and geometry. It can be generalized to cases with an arbitrary number of reservoirs attached to an arbitrary number of sites in $\qs$. 

On reaching the NESS, $\cm(p\tau+\tau)=\cm(p \tau)$. This makes Eq.~\eqref{eq:ARC} exactly of the form of a discrete--time Lyapunov equation (an equation widely studied in mathematics and engineering). The discrete--time Lyapunov equation can be solved directly, allowing to find the NESS without carrying out the time evolution. Moreover, the timescale, or the number of cycles, for reaching the NESS in an ARC process is given by the largest magnitude eigenvalue of $\mathbf{M}$. Indeed, following the standard stability analysis of the dynamical process in Eq.~\eqref{eq:ARC}, the timescale to convergence to the NESS is
\[ \label{eq:TC}
\timeNESS = \frac{ \tau} { -\ln(|m_0|)},
\]
where the largest magnitude eigenvalue, $m_0$, of $\mathbf{M}$ dictates the number of cycles and $\tau$ is the unitary evolution for a single cycle.  When all eigenvalues have magnitude less than one, the steady state is unique. In our example, the largest eigenvalue is such that $|m_0| < 1$ and uniqueness is guaranteed. Equation~\eqref{eq:TC} is something we will employ to estimate one of the contributions to the computational cost in tensor networks, where explicit time evolution is currently needed to obtain the NESS. 

Finally, given the correlation matrix of $\ql \qs \qr$ at any time (or in the NESS), the particle current 
\[ \label{eq:Curr}
I_{\ql \qs}=2 \sum_{k\in\ql,i\in\qs} \Im v_{ik} \cm_{ki}
\]
is flowing across the $\ql \qs$ interface (assuming that $\ql$ is only connected to $\qs$ and not directly to $\qr$). Similar expressions apply to other interfaces.

\section{Results}
\label{sec:results_arc}

The primary goal of this work is to extensively characterize ARC and its limits (CR and PR) in order to understand their behavior and performance (error--cost relationship), and determine the optimal approach within the confines of the example system. We will employ the approach of Sec.~\ref{sec:NonintARC} to find the NESS, including its current, in order to compare the results with the reference (continuum, macroscopic) results of Sec.~\ref{sec:RefResults}. Before proceeding to the NESS properties, we first examine the full time evolution within ARC, which will elucidate the underlying processes within this framework.

\begin{figure}[t!]
\includegraphics[width=\columnwidth]{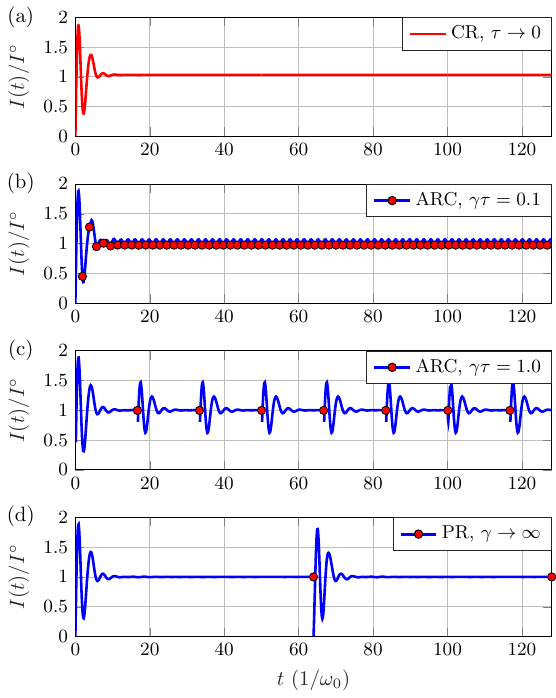}
\caption{{\bf Time dynamics of ARC.} The plots show the current between $\ql$ and $\qs$ versus time for $\NW=128$ sites per reservoir. The ARC parameters are (a) $\tau\to 0$ at finite $\gamma \approx 0.084 \, \omega_0$ (i.e., in the CR regime), (b) $\gamma\tau=0.1$ and $\tilde{\tau} \approx 37.90 / \omega_0$, (c) $\gamma\tau=1$ and $\tilde{\tau} \approx 37.30 / \omega_0$, and (d) $\gamma\to\infty$ at finite $\tau = 64 / \omega_0$ (i.e., in the PR regime). The red points indicate the current right before the application of dissipation, whereas CR is a continuous red line since relaxation is continuous. For all regimes of ARC except the CR limit, the stroboscopic points in red are the most relevant. The convergence of these at long times gives the estimate of the current within the exact NESS. We note that (a) $\gamma$ for CR is provided by the heuristic procedure of Ref.~\onlinecite{elenewski_performance_2021}, [(b) and (c)] $\action$ is given by the value that best estimates the current at fixed $\gamma\tau$, and (d) $\tau$ is given by the (physical) refresh time for PR, $\tau = \TW = \NW/2 \omega_0$ with $2 \omega_0$ the Fermi velocity.}
\label{fig:3}
\end{figure}

\subsection{Time dynamics}

Figure~\ref{fig:3} plots the (normalized) current $I(t)/\Io$ between the system and the left reservoir for the resonant level model (Sec.~\ref{sec:model} and Fig.~\ref{fig:2}). Four representative ARC parameter regimes are in the figure. In the CR limit ($\tau\to 0$ at finite $\gamma$), Fig.~\ref{fig:3}(a), there are initial oscillations but the current ultimately relaxes close to the reference value $\Io$, as seen by the approach of the normalized current to 1. Away from this limit, the cyclic nature of the ARC becomes apparent. In Fig.~\ref{fig:3}(b,c), the intermediate values of $\gamma\tau$ show the initial oscillations, as well as oscillations after each partial refresh. The NESS defined by the periodic stroboscopic points in red shows that the current still relaxes near to $\Io$. Figure~\ref{fig:3}(d) provides the PR limit ($\gamma\to\infty$ at finite $\tau$). This also has oscillations but, at the end of each cycle, the current at the $\ql\qs$ interface goes to zero since the correlations between the system and reservoirs are completely removed. As with the other cases (except the CR limit), the stroboscopic points define the NESS and converge to $\Io$.

\subsection{Phase diagram for the steady state}

\begin{figure}[t!]
    \centering
    \includegraphics[width=\columnwidth]{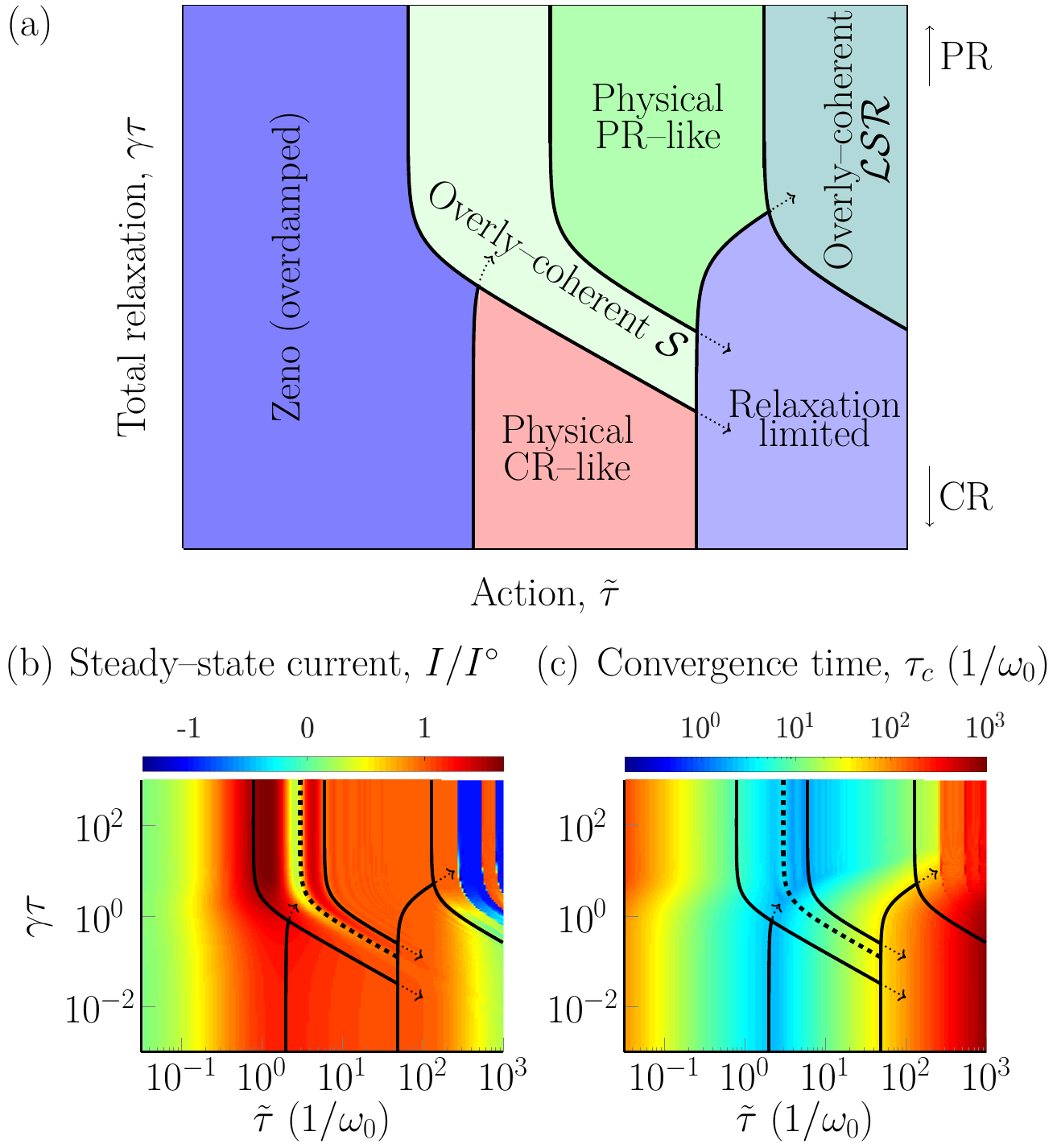}
    \caption{{\bf Phase diagram of ARC.} (a) The schematic phase diagram (of the NESS current) shows several regimes as the total relaxation $\gamma\tau$ and the action $\action = |\tau~-~\im 2/\gamma|$ vary at fixed $\NW$. In this parameterization, the PR limit ($\gamma \to \infty$ at fixed $\tau$) is on the upper boundary and the CR limit ($\tau \to 0$ at fixed $\gamma$) on the lower. Sweeping the action from small to large values shows three general regimes. On the left is an overdamped (Zeno) regime (i.e., large $\gamma$ and small $\tau$ corresponding to a small action). On the right is an underdamped regime. The weak relaxation regime on the bottom right (i.e., with CR--like behavior) leads to currents limited by the relaxation strength. In the overly coherent regime in the top right (i.e., with PR--like behavior), there are oscillations across the $\ql \qs \qr$ system. The middle of the diagram contains the physical regimes, with a CR--like physical regime at the bottom and PR--like at the top. The physical regimes are separated by a region of an overly coherent $\qs$. (b) The normalized current for $\NW=256$ on the same diagram. The lines of constant $\gamma$---the ones coming vertically from the bottom---are $\omega_0/43.3$ and $\omega_0$ which are heuristic optimal for CR and transition to overdamped regime for CR, respectively. 
    These are normal turnover points discussed extensively elsewhere~\cite{gruss_landauers_2016,elenewski_communication_2017,wojtowicz_open-system_2020,wojtowicz_dual_2021}. The solid lines of constant $\tau$---the ones coming vertically form the top---are from left to right, $\tau^\star = \pi/\qw$, $2 \TS$, and $\TW$. The former two define the transition regime. {The dashed line is $\TS$, showing it also demarcates features.} The $\TW$ shows the transition to an overly coherent $\ql\qs\qr$. To the right of $\TW$, the current remains a good approximation until $2\TW$ since that is when the density wavefront returns to the system after bouncing off the finite reservoir boundary. (c) The convergence time to NESS follows from Eq.~\eqref{eq:TC}.
    For another perspective on the phase diagram, Fig.~\ref{fig:s5} shows it within the $\gamma-\tau$ parameterization.
    }
  \label{fig:4}
\end{figure}

The steady state of ARC---specifically, the level of current that its NESS sustains---yields different phases of behavior versus the total relaxation $\gamma\tau$ and action $\action = |\tau~-~\im 2/\gamma|$, see Fig.~\ref{fig:4}(a). These phases are analogous to the Kramers' turnover regimes discussed extensively in the context of ERA with continuous relaxation for quantum electron~\cite{gruss_landauers_2016,elenewski_communication_2017,wojtowicz_open-system_2020,wojtowicz_dual_2021} and classical thermal~\cite{velizhanin_crossover_2015,chien_thermal_2017} transport. In Sec.~\ref{sec:cr}, we introduced these phases for CR. There are relaxation--limited currents for small $\gamma$ and Zeno--limited currents for large $\gamma$. In between, there is an intermediate--$\gamma$ regime where the intrinsic, physical current dictates the transport behavior. This turnover is shown versus action $\action$ (which depends on the inverse of $\gamma$) in Fig.~\ref{fig:4}(a), so, from left to right, the behavior goes from Zeno through the physical regime to the relaxation--limited regime. When the behavior is more like PR, there is a Zeno regime that goes to a physical regime, but the relaxation--limited regime is now characterized by overly coherent oscillations in the global $\ql\qs\qr$ system. 

The Zeno behavior for small action occurs when $\action$ is shorter than any timescale in $\ql\qs\qr$. On the PR side, this corresponds to frequent refreshes, i.e., small $\tau$, where repeated resets of the reservoirs occur before the current fully starts to flow (i.e., before the restoration of sufficient coherence between the system and reservoirs). This is analogous to the quantum Zeno effect in frequently measured systems~\cite{misra_zenos_1977}. For CR, the physics is similar, but now the continuous (or quasicontinuous) relaxation is strong, giving rise to the same effect. Eventually, the current will steadily decrease to zero as the action decreases (i.e., as $\tau$ decreases and $\gamma$ increases) in this regime.

In the underdamped regime, unlike the Zeno regime, there is distinct PR--like and CR--like behavior. In the former, there are global oscillations of the current as its wavefront transverses $\ql\qs\qr$ and bounces off the finite size boundaries. The time for seeing the effect of the boundary is related to the finite speed of propagation given by the Lieb--Robinson bound. An unmistakable signature for this over--coherence is the inversion of the current when $\tau=2\TW$. At twice the physical timescale $\TW$, the density wavefront initiated at the time of refresh will travel to the boundary and back to the system in one cycle, reversing the current~\cite{chien_landauer_2014}. Further increasing $\tau$ by $2\TW$ will lead to other reversals, and so on. For CR--like behavior in the underdamped regime, such oscillations do occur and influence the character of the steady state. Yet, due to the continuous or quasicontinuous character of the relaxation, the current becomes simply linearly proportional to the relaxation strength, as the contact to the external environments is the dominant resistance. 

The physically relevant regime of the current occurs at intermediate $\action$. The character of intermediate phase depends on $\tau$ and $\gamma$, as the ARC cycle is the first order Trotter decomposition of the evolution superoperator generated by Eq.~\eqref{eq:evolution_cr}. Ignoring $\gamma$ for a moment, the decomposition is exact for $\tau\rightarrow 0$ but we can still get reasonable agreement for a finite $\tau$, i.e., quasicontinuous behavior. However, if the coherent evolution is long enough that substantial changes in $\ql\qs\qr$ take place during the evolution $\tau$, then the Trotterization will no longer approximate the continuous dynamics. The smallest important timescale will   depend on the model. We will consider  
\[ \label{eq:risetime}
\tau^\star=\pi/\qw,
\]
which is the rise time for the formation of the steady state when the reservoir has  bandwidth $\qw$~\cite{zwolak_communication_2018}. This time is usually smaller than other relevant timescales in the $\ql\qs\qr$ model as the bandwidth is large. In the PR limit, the time $\tau^\star$ should bound the Zeno regime, as the current forms at the interface in this time. Moreover, for $\tau > \tau^\star$, CR--like behavior breaks down, yet $\ql\qs\qr$ still does not have fully formed global coherence. That coherence requires $\tau$ to be greater than the time it takes for particles to transit the system. Similar to $\TW$ for the reservoirs, Lieb--Robinson arguments suggest a transit time of
\[ \label{eq:TS}
\TS = \NS / 2v_\qs ,
\]
where $2v_\qs$ is the (effective) Fermi velocity in the system. For the parameters of our example, $\NS = 3$ and $v_\qs=\omega_0 /2$, this is $\TS = 3/\omega_0$. For $\tau < \TS$, the system $\qs$ accepts some particles from the reservoirs, but these are internally overly coherent (i.e., they oscillate within the system but have their coherence with the reservoirs cutoff before $\ql \qs \qr$ coherence forms). Similarly to $\ql$ and $\qr$, for $\tau$ multiples of $\TS$, we will see the reflection of particle front traveling from one interface to another. 

All parts of a parameter space where quasicontinuity breaks down will have refresh--like physics. However, this regime of overly coherent $\qs$ has a distinct nature, as seen in Fig.~\ref{fig:4}(b). This regime extends between $\tau^\star$ and $2\TS$. These two times demarcate the transition region between physical CR--like and PR--like regimes. We note, though, that the upper limit on $\tau$ is not a hard transition. In principle, one should have features at higher multiples of $\TS$. Yet, their visibility will depend on details of the model, such as the probability to transmit across the system--reservoir interfaces (which depend on the system--reservoir coupling). Our example  sustains a relatively high current and many reflections within $\qs$ only give small corrections. Later we will introduce an averaging procedure to eliminate this timescale in determining the performance of ARC (see Sec.~\ref{sec:convergence}).

To summarize, at fixed, but not too large, $\gamma$, ARC remains quasicontinuous so long as $\tau$ is smaller than $\tau^\star$. On the diagram, this means following the lines of fixed $\gamma$, which start vertically from the bottom but then turn to the right. When these hit the lines of constant $\tau = \tau^\star$, the behavior goes through a transition region---the overly coherent $\qs$ region. Increasing $\tau$ beyond about $2 \TS$ (for our example) stabilizes the current.

The physics within this diagram is further illustrated by the bottom panels of Fig.~\ref{fig:4}. Figure~\ref{fig:4}(b) shows the normalized current. This is the quantity that defines the phase diagram, showing all the regimes and features. The meaning of signature lines in $\gamma$ and $\tau$ when drawing the phase diagram in terms of these parameters is in App.~\ref{si:arc_phasediagram2}. 
Figure~\ref{fig:4}(c) shows the timescale to reach the steady state (derived from $\mathbf{M}$). This timescale also has signatures of the lines of constant $\tau$ and $\gamma$, but does not show all the regimes that are seen in the phase diagram. Additional information on ARC is included in App.~\ref{si:arc_growNSNW}.

\subsection{Convergence}
\label{sec:convergence}

Now that we have established the character of ARC, we can examine the error--cost relationship. We will characterize error in terms of two quantities. One is in terms of the steady--state current, which is often the main---or only---quantity of interest. The other will be in terms of the system's correlation matrix. This is error--``agnostic'' and captures how well one can reproduce arbitrary observables on the system (e.g., densities, etc.). 

For the current, we will focus on the interfacial currents between $\ql$ and $\qs$, $I_{\ql\qs}$, and between $\qs$ and $\qr$,  $I_{\qs\qr}$. Specifically, we will have the method (ARC with a particular set of parameters) output a current estimate
\[ \label{eq:CurrEst}
I = (I_{\ql\qs}+I_{\qs\qr})/2 ,
\]
which is an average current (across the two interfaces in this case; one could take this across more partitions). The estimate of the relative error for this current, $\sig^2$, will be a combination of two independent sources that contribute to the loss of accuracy and precision, 
\[ \label{eq:CurrErr}
\sig^2 = \sigma_1^2+\sigma_2^2 .
\]
The first contribution is just the relative accuracy, 
\[ \label{eq:sig1}
\sigma_1^2 = \left( \frac{I-\Io }{\Io} \right)^2 .
\]
It is tempting to stop there when one has the exact result, $\Io$, to compare to. However, except for the exact solution of the CR limit, one can have cases where there are mismatched currents at the interfaces, $I_{\ql\qs} \neq I_{\qs\qr}$, that are used to provide the current estimator, Eq.~\eqref{eq:CurrEst}. One can thus obtain the exact current but still have some underlying error where the current is not reflecting a proper stationary state (i.e., $I_{\ql\qs}=I_{\qs\qr}$). In PR, the mismatch can occur because  $\ql\qs\qr$ is still evolving right before the refresh. The existence of a NESS is only at the level of the state, but not at the level of the currents flowing at different partitions. Alternatively, if one does not have the exact solution but rather one uses tensor networks, the truncation of the state can make currents not match~\cite{wojtowicz_open-system_2020}. The second contribution is thus the relative mismatch,
\[ \label{eq:sig2}
\sigma_2^2 = \frac12 \left(\frac{I-I_{\ql\qs}}{\Io}\right)^2 + \frac12 \left(\frac{I-I_{\qs\qr}}{\Io}\right)^2  = \left(\frac{I_{\ql\qs}-I_{\qs\qr}}{2 \Io}\right)^2,
\]
which provides a measure of precision either due to numerical errors in calculations or due to refresh--induced properties of the NESS. 

However, the above contributions still do not properly account for accidental crossings that can happen, e.g., when $I$ is oscillating in the parameter space around the thermodynamic-limit value $\Io$. This is indeed the case in PR--like phase and at an accidental crossings for small and large $\gamma$ in the Kramers' turnover in the CR limit~\cite{gruss_landauers_2016,elenewski_performance_2021,wojtowicz_dual_2021}. Those are times when both the contributions are zero, $\sigma_1=\sigma_2=0$, but the current is still not properly reflecting the continuum limit result.

For that reason, we quantify convergence by calculating a mean relative error,
\[ \label{eq:sigmean}
\asigsq(\action) = \mathrm{mean}_{|\action^\p - \action| \le \TS/2} [ \sig^2(\action^\p) ]
\]
at fixed $\gamma\tau$. A moving average smooths the outcome over a window of the size $\TS$ using Eq.~\eqref{eq:TS}, that corresponds to the time period of the natural oscillation in the full refresh limit due to the system, and this averaging removes this effect (and associated accidental crossings). Consequently, $\asig$ gives us a measure of consistency. An extracted current is only considered good if neighboring values of parameters are also giving good estimates. Thus, accidental crossing will not be assigned zero error. Additional information on this procedure is in App.~\ref{si:arc_averaging}.

\begin{figure*}[t!]
    \centering
    \includegraphics[width=\textwidth]{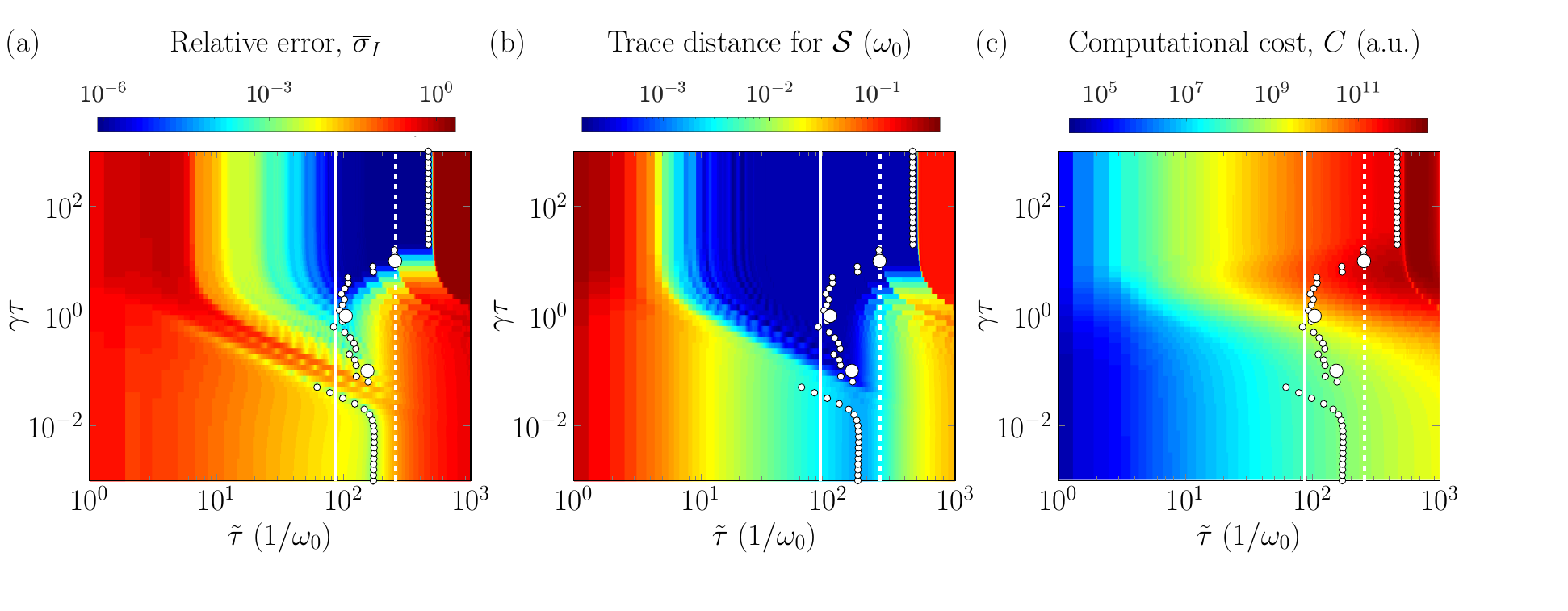}
    \caption{{\bf Convergence and computational cost of ARC.} The three panels show the (a) current error, Eq.~\eqref{eq:sigmean}, (b) the correlation matrix error, Eq.~\eqref{eq:CorrErr}, and (c) the estimated computational cost, Eq.~\eqref{eq:EstCompCost}, for our example system and for $\NW = 512$, $\beta\omega_0=40$, and $\mu=\omega_0/2$. The two error plots have many of the same features as the phase diagram for ARC, Fig.~\ref{fig:4}, as we expect since the different regimes directly impact the error. Using the current error as the quantity to be optimized, the best choice of action for a fixed $\gamma\tau$ is marked by the white points in (a). These points are replicated in the other plots, [(b) and (c)], to see where they fall for other quantities. These points are close to where the correlation matrix error is minimum in (b), but they do not follow any features in the computational cost in (c). As is evident, CR-- and PR--like regimes have very different estimated computational costs. Within the physical regimes, this cost is correlated with the error, with cost going up for more accurate simulations. Outside of these regimes, though, there is no correlation. The whites dots for $\gamma\tau=0.1, \, 1$ and $10$ are made thicker to guide the readers eye. Vertical solid line represents $\action$ equal to heuristic choice for CR, and vertical dashed line represents $\action=\TW$, which is the physical choice for PR. We employ these $\action$ for the CR and PR limits.
    }
    \label{fig:5}
\end{figure*}

Figure~\ref{fig:5}(a) shows how the mean relative error $\asig$ behaves within the parameter space of ARC. The features present reflect the phase diagram for ARC, with the lowest relative error for physical CR--like and PR--like phases. At fixed $\NW$, the physical PR--like phase generally performs better then physical CR--like phase due to capturing more of the non--Markovian character of the evolution. In addition, the PR--like phase has quite a large plateau, a region with low error versus $\gamma \tau$ and $\action$. This is a consequence of the rapid rise of the current to the steady state in closed quantum systems~\cite{zwolak_communication_2018} (i.e., the coherent evolution) together with the stabilization provided by the periodic refresh of the reservoirs. Within the figure, we also show white points. These are the points of lowest $\asig$ for a fixed $\gamma \tau$. They represent what action should be taken for a given $\gamma \tau$ in order to get the best estimate of the current. In order to compare ARC at different $\gamma\tau$, we will use the action at these points. 

The convergence of the correlation matrix for $\qs$ largely follows the convergence of the current, see Fig.~\ref{fig:5}(b), which shows the trace distance to the reference state, 
\[ \label{eq:CorrErr}
\frac{1}{2}  \tr \left| \cm_\qs - \CSo \right| .
\]
Just like the current error, the plot has many of the same features as the phase diagram for ARC. There is a clear cut between physical CR--like and physical PR--like with PR--like having notably smaller error than CR--like. There is a Zeno regime on the left and underdamped regimes on the right (that split into the overly coherent $\ql\qs\qr$ and relaxation--limited regimes, for the PR and CR sides, respectively). The site occupancies (i.e., the diagonal elements in the position basis) dominate the correlation matrix error of $\qs$, as the current flow is only a small contribution to the total correlation matrix~\cite{elenewski_performance_2021}. This is model dependent. A setup that carries a larger current may have current contributions to the error of a similar magnitude. However, we always have a bound $|\cm_{mn}| \le \sqrt{\cm_{mm} \cm_{nn}}$. 
This, in turn, bounds the current by the product of onsite occupations, $\cm_{mm}$ and $\cm_{nn}$. 
Thus, the current error can at most be similar in magnitude to density errors, but it can not dominate. As a consequence of the outsized importance of occupations, the trace distance does not contain all the same subtle features as the current error. Figure~\ref{fig:5}(b) also shows that the points of smallest $\asig$ (the white points) match with the regions of good convergence for the correlation matrix.

\subsection{Estimated computational cost}
\label{sec:cost}

Figures~\ref{fig:5}(a) and ~\ref{fig:5}(b) both demonstrate that physical PR and PR--like regimes of ARC provide highly converged results, whether quantified by the current alone or the correlation matrix, at a fixed reservoir size. The results are much more accurate than the physical CR and CR--like regimes for the same reservoir size. Ultimately, however, one needs the level of error for a given computational cost and how the error scales with computational cost within some many--body method of choice. 

In this section, we will estimate the computational cost, $\CC$, for a particular tensor network implementation (described in the final subsection before the conclusions) via quantities computed with non--interacting simulations. In the next two subsections, we examine the error--cost scaling and the convergence within actual tensor network simulations. In particular, we exploit the non--interacting nature of our example and extract relevant quantities from the various tools presented in this work. The relevant quantities will be the time to reach the steady state, the lattice size (of order $\NW$), and the operator space entanglement entropy (OSEE)~\cite{prosen_operator_2007, dubail_entanglement_2017}, $\OSEE$, with the latter enabling an estimate of the matrix product dimension $D$. We summarize the procedure to calculate $\OSEE$ from the correlation matrix $\cm$ in App.~\ref{si:osee}. 

For the tensor network techniques, we employ (see the final subsection before the conclusions), the estimated computational cost is
\[ \label{eq:EstCompCost}
\CC \sim \timeNESS \cdot \NW  \cdot 2^{3 \cdot \OSEE} .
\]
For the last term, we use binary entropy to define $\OSEE$, leading to base 2 exponentiation, and calculate it in the middle of the lattice as an approximation.  When employing tensor networks directly, the cost is taken as
\[ \label{eq:EstCompCostMPS}
\CC^\mathrm{MPS} \sim \timeNESS \cdot \NW  \cdot D^3 .
\]
We will explain each of these pieces in turn below.

Currently, no tensor network technique has been demonstrated to directly target the steady state for the class of problems we consider. Rather, real--time runs evolve the state until the state is stationary up to some tolerance. As introduced earlier, the quantity $\timeNESS$, Eq.~\eqref{eq:TC}, from the map $\mathbf{M}$ determines the convergence. A real--time run on the order of this timescale will exponentially suppress all contributions to the state except the NESS. The combination of the first two relevant factors, $\timeNESS \NW$, thus gives the size of the problem. The tensor network simulation will have to iterate over a number of steps proportional to $\timeNESS$ times a number of lattice sites proportional to $\NW$. The latter follows from the observation that the relevant entanglement appears within the whole bias window, which contains a number of modes proportional to $\NW$.

\begin{figure}[b!]
    \centering
    \includegraphics[width=0.8\columnwidth]{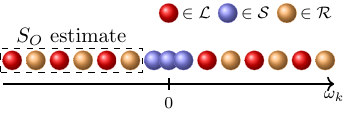}
    \caption{\textbf{Mixed basis.} For MPS simulations and the calculation of OSEE, we group the modes from reservoirs $\ql$ and $\qr$, in the frequency basis, and ordered them according to $\omega_k$ in Eq.~\eqref{eq:hlr}. It reflects the emergent structure of correlations in the reservoirs~\cite{rams_breaking_2020}. We place the system modes around zero frequency. We indicate the bipartition of the lattice for calculating OSEE. }
    \label{fig:6}
\end{figure}

The remaining factor is the exponent of $\OSEE$. With matrix product states (MPS), the convergence of the state is determined by the matrix product dimension, $D$ (other parameters, such as the Schmidt tolerance, can control convergence, but ultimately these just determine $D$). The time evolution within the implementation we employ (see Sec.~\ref{sec:error_cost}) has a scaling of the computational cost with $D^3$, Eq.~\eqref{eq:EstCompCostMPS}. Non--interacting simulations can not provide the exact $D$. Moreover, they can't tell us how error will depend on $D$. However, one can estimate the $D$ necessary for convergence (to the exact finite $\NW$ result) as $D \sim 2^{\OSEE}$. This provides the last piece in the estimate in Eq.~\eqref{eq:EstCompCost}. We emphasize that the cost estimates are intended to capture the scaling. There will be proportionality factors including, e.g., the inverse of the time step and the local Hilbert space dimension to the third power. Both are not changed in our case and thus they are left out.

We also note that we employ what is known as the {\em mixed basis} for the tensor network~\cite{rams_breaking_2020}, see Fig.~\ref{fig:6}. The choice (and order) of basis has an important impact on the structure of correlations within the network that, in turn, influences the required bond dimension $D$. In the mixed basis, we use the frequency (or, in higher dimensions, momentum) basis and jointly order the $\ql\qr$ modes according to their frequencies. This reflects natural scattering structure for elastic transport where a particle of frequency $\omega_k$ scatters from $\qs$ and forms an entangled pair shared between modes of that frequency in $\ql$ and $\qr$. Placing these modes next to each other make the entangled pair localized in the MPS description, which in turn lowers the overall cost. In our implementation, the reservoirs' modes are ordered by increasing frequencies and $\qs$ is placed in the middle of bias window. The placement of $\qs$ is to reduce the spread of correlations across the MPS by placing $\qs$ close to the modes that contribute to the current. This construction drastically reduces the $\OSEE$ compared to working, e.g., in the spatial basis (i.e., for our case, a one--dimensional nearest--neighbor hopping model).

Figure~\ref{fig:5}(c) shows the estimated computational cost for the mixed basis. ARC has the lowest cost (with no consideration of errors) when the parameters specify frequent and strong relaxation. That relaxation brings the MPS towards a state that is a product over all reservoir modes. Veering toward this state lowers the bond dimension. However, the parameter regimes where the state is close to a product state (for the reservoirs) is not representative of the physics that one is trying to simulate with ERAs. Rather, one needs correlations to build up between the system and reservoirs. This creates a trade-off between improving convergence and reducing computational cost. From Fig.~\ref{fig:5}, while not one--to--one, there is clearly a relation between cost and accuracy. Since we always work with limited computational resources, we would like to know how the error--cost relationship scales in different ARC regimes and whether there is an ideal regime to work in.

\subsection{Error--cost scaling}\label{sec:error_cost}

\begin{figure}[t!]
    \centering
    \includegraphics[width=0.9\columnwidth]{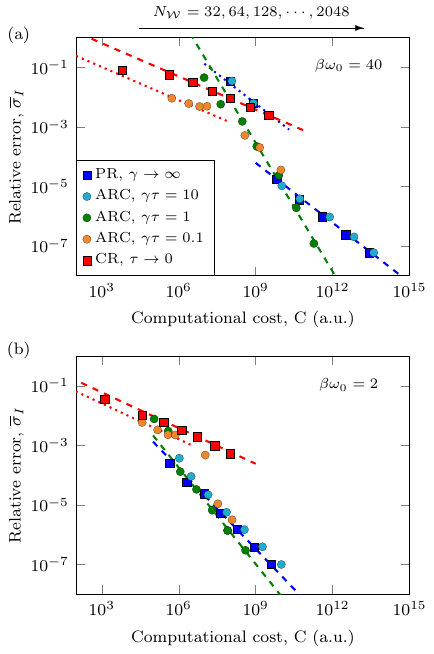}
    \caption{\textbf{Error--cost scaling for ARC.} The error is shown versus the estimated computational cost, Eq.~\eqref{eq:EstCompCost}, for (a) a low and (b) a high temperature both with bias $\mu=\omega_0/2$. For a given $\gamma\tau$, the reservoir size $\NW$ increases from left to right (as indicated on the top of the figure), which is the only control parameter here since the estimated computational cost comes from non--interacting simulations. For ARC, the action is taken at the white points in Fig.~\ref{fig:5}. For CR, we choose the relaxation from heuristic approach~\cite{elenewski_performance_2021}. For PR, we choose $\tau=\TW$. While there are some transitions in the error--cost scaling, most of the data settles into a power law behavior, Eq.~\eqref{eq:ErrScaling}. The dashed lines show fitted scaling for CR (red), ARC $\gamma\tau=1$ (green), and PR (blue). Dotted lines mark related scaling laws. These are not fits but a reference lines of the same $\nu$ as their dashed counterparts. More information on the fit can be found in Sec.~\ref{sec:error_cost} and App.~\ref{si:arc_scaling}.
    }
    \label{fig:7}
\end{figure}

As we discuss above, CR and PR methods converge to the continuum, macroscopic reservoir limit as $\NW\rightarrow\infty$ (with $\gamma \to 0$ or $\tau=\TW$, respectively). However, we want to know how the error scales with increasing $\NW$. Figure~\ref{fig:7} compares the error--cost scalings for CR, PR, and other ARC approaches at fixed $\gamma\tau$. Here, we take CR at the heuristic point, see Sec.~\ref{sec:cr}. For PR, we use the physical action $\tilde\tau=\TW$, as described in  Sec.~\ref{sec:pr}. ARC regimes between these two employ the $\action$ that gives the lowest current error $\asig$ (the white points in Fig.~\ref{fig:5} and equivalent points for other $\NW$). For each $\gamma\tau$, the reservoir size $\NW$ increases from left to right in Fig.~\ref{fig:7}. 

The error--cost scaling within the physical regimes nicely separates into three classes of behavior: physical CR--like, intermediate, and physical PR--like (the latter are both on the PR--like side of the cut in the phase diagram). The whole physical CR--like regime behaves similarly to the CR limit. We observe varying scaling in the physical PR--like regime. One is for physical PR--like, which includes the PR as a limiting case. Another for the intermediate relaxation regime that represents a case lying in the middle of the phase diagram close to the transition from PR--like regime to the overly--coherent $\qs$ regime. Here, the advantages of CR and PR reinforce each other, resulting in favorable error--cost scaling. The intermediate regime has both moderate OSEE (lowering the cost) and sufficient non--Markovianity (increasing accuracy). 

Note that the borders of those three regimes follow from the phase diagram and are dependent on $\NW$. The scaling can transition from one relation to another as $\NW$ grows. This transition can be seen for $\gamma\tau=0.1$, which exhibits a scaling characteristic for the physical CR--like regime for $\NW\leq 256$, and switches to the intermediate--relaxation scaling for larger $\NW$. The transition happens when the optimal action (the white points in the figure) jumps from the CR--like side to PR--like side, compare Fig.~\ref{fig:s7} with Fig.~\ref{fig:5}. 

In Fig.~\ref{fig:7}, we show the scaling relation at low temperature $\beta\omega_0=40$, Fig.~\ref{fig:7}(a), and at high temperature $\beta\omega_0=2$, Fig.~\ref{fig:7}(b). In both plots, we determine the scaling relations using CR for the physical CR--like regime, PR for PR--like, and ARC of $\gamma\tau = 1$ for the intermediate case. We perform fits to $\NW\geq 128$, to best capture the scaling limit.

The error of the current is typically smallest in the physical PR--like regime, which also leads to the accumulation of more non--Markovian character. However, it is the intermediate regime that has the best proportionality in terms of $\NW$. For the physical PR--like regime, the error of the current goes down approximately as $\NW^{-2}$ for both low and high temperatures. In the intermediate relaxation regime, we have scaling $\NW^{-3.4 \pm 0.2}$ and $\NW^{-2.21 \pm 0.04}$ for low and high temperature, respectively. For physical CR--like, the error of the current goes down as $\NW^{-0.91\pm 0.02}$ and $\NW^{-0.87\pm 0.02}$ for low and high temperatures, respectively. 

The scaling of the computational cost goes down as the temperature increases. This is driven by the scaling of $2^{\OSEE}$. In particular, its exponents drop by a factor of 2 to 3 times when going from low to high temperature. The decrease reflects that in the limit of infinite temperature, the MPS representation would correspond to a product state with zero OSEE. Separately, the OSEE grows fastest in the physical PR--like regime, where coherent dynamics create a lot of entanglement, and the slowest for the physical CR--like regime. The other contribution to the cost, the convergence time $\timeNESS$, grows similarly for high and low temperature, being proportional to $\NW$ for physical CR--like, $\NW^{0.8}$ for physical PR--like, and $\NW^{0.66\pm 0.04}$ to $\NW^{0.80\pm 0.01}$ for low and high temperature for the intermediate case, see App.~\ref{si:arc_scaling} for details. 

We fit the error--cost relationship using the form
\[ \label{eq:ErrScaling}
\asig = A C^{-\nu} ,
\]
where $A$ is a coefficient, $\nu$ is the scaling exponent of the decay, and $\CC$ is the computational cost. Note that the contributions to $\nu$ can be delineated by looking at scaling of each component in Eq.~\eqref{eq:EstCompCost} with $\NW$, see App.~\ref{si:arc_scaling}. In Fig.~\ref{fig:7}, we mark the fits by dashed lines and use dotted lines to indicate the same scaling exponent within relevant regions.  At lower temperature, $\beta\omega_0=40$, the fits give $(\log(A),\nu)$ equal $(4.2 \pm 0.3, 0.67 \pm 0.01)$, $(21.7 \pm 1.5; 1.44 \pm 0.07)$, and  $(2.14 \pm 0.06, 0.372 \pm 0.003)$ for PR, $\gamma\tau=1$, and CR, respectively. The dotted lines show the same scaling exponent as the corresponding dashed line (of the same color) but with a different coefficient $A$. The red dotted line underlines the scaling relation of $\gamma\tau=0.1$ before it transitions to intermediate ARC scaling. This stresses that it undergoes the same scaling relation as for CR--limit at heuristic $\gamma$. The dotted line is lower than for CR itself because the optimization procedure targets a different action than the CR heuristic. The blue dotted line provides the scaling relation of PR before $\NW$ exceeds the thermal correlation length $\NT$, after which the prefactor of the fit decreases substantially. We discuss this further below. The error--cost scaling is best for $\gamma\tau=1$ and that PR outperforms CR by about a factor of two for the exponent. Nevertheless, the prefactor is important for achieving a given error when there are constraints on available computational resources. The higher level of correlations retained in the PR approach, in particular, makes it prohibitive at low temperatures within our tensor network approach at the present level of resources. 

At high temperatures, Fig.~\ref{fig:7}(b), the qualitative observations on performance carry over. However, the intermediate and PR regimes have much smaller prefactors $A$, with the performance of those regimes crossing the CR at smaller $\NW$. At $\beta\omega_0=2$, the pairs $(\log(A),\nu)$ are $(3.7\pm 0.4; 0.90\pm 0.02)$, $(6.3\pm 0.4; 1.08 \pm  0.02)$, and $(0.0\pm 0.2; 0.40 \pm 0.01)$, for PR, $\gamma\tau=1$, and CR, respectively. The red dotted line underlines the scaling relation of $\gamma\tau=0.1$ before transitioning to intermediate scaling. 

\begin{figure}[h!]
    \centering
    \includegraphics[width=\columnwidth]{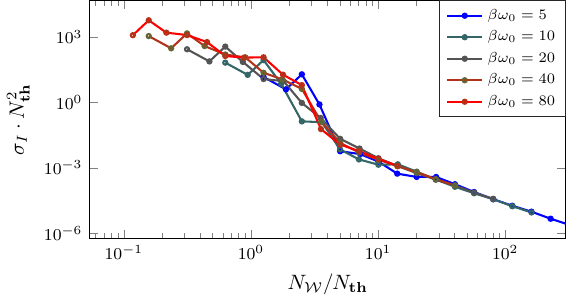}
    \caption{{\bf Thermal correlation length in the PR regime.} We show the relative current error rescaled by $\NT^2$ versus the number of reservoir modes divided by $\NT$. We observe the collapse of the data for a range of temperatures onto a single curve, that exhibits two scaling regimes, one for $\NW \lesssim \NT$ and the other for $\NW \gg \NT$. In both regimes, the error vanishes as $\NW^{-2}$, but with different prefactors. }
    \label{fig:8}
\end{figure}

Finally, we already have seen the role of $\tau^\star$ and $\TS$ is determining the transition into the physical PR--like regime, and thus their influence on errors. There is also a sudden drop in performance in the PR limit when going from smaller $\NW$ to larger $\NW$ in Fig.~\ref{fig:7}(a). This is due to another length (time) scale, the thermal correlation length for the reservoirs.  For a 1D uniform lattice reservoir, the thermal correlation length, calculated in isolation from the system and at half-filling, is 
\[
\NT=2\beta\omega_0/\pi,
\label{eq:nth}
\]
{where, again, $\beta=\hbar/k_BT$ is the thermal relaxation time for the reservoir}. This has an associated timescale
\[
\TT=\NT/2\omega_0 ,
\]
which is defined similarly to $\TW$. In Fig.~\ref{fig:8}, we focus on the PR limit and show the decay of the relative current error $\sig$ with the number of reservoir's modes $\NW$ for a wide range of temperatures. The data collapse on a single curve when $\NW$ is rescaled by $\NT$ and the error multiplied by $\NT^2$. The curve features a characteristic drop. It is important to note that there are also other sources of error, such as the one related to $\qs$, which can also result in a similar drop. Nevertheless, Fig.~\ref{fig:8} illustrates that for the best performance one may require $\NW \gg \NT$ in the PR limit. This is identical to a  consideration for CR, which requires that {$\beta \gamma \le 1$} and yields an associated bound on the reservoir size $\NW \gtrsim \beta\omega_0$ (in the absence of other relevant factors that can mitigate the effect)~\cite{gruss_landauers_2016,elenewski_communication_2017}.

\subsection{Tensor network simulations}

\begin{figure}[t!]
    \centering
    \includegraphics[width=\columnwidth]{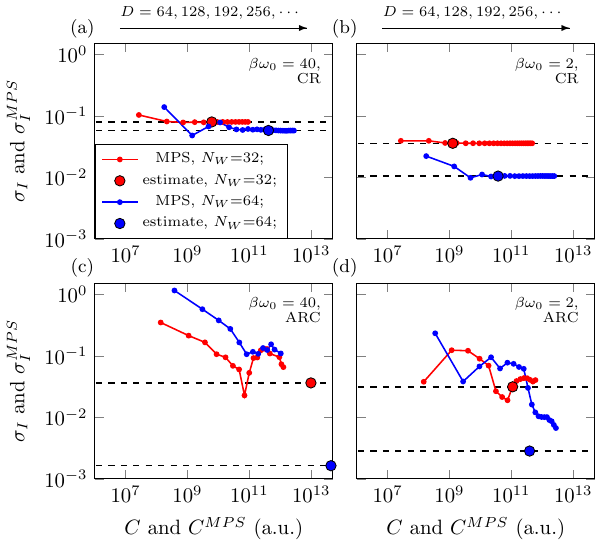}
    \caption{{\bf Convergence with bond dimension.} The plots show the error versus cost obtained using the MPS implementation as described in the main text. The data were obtained for (a) CR and (b) ARC $\gamma\tau=1$ at low and high temperature as indicated on the panels. The matrix product dimension, $D$, is increased at fixed $\NW$ (shown in the figure), starting from $D=64$ (at the left) and increasing by $64$ between consecutive points. While technically still an estimated computational cost, $\CC^\mathrm{MPS}=\timeNESS \cdot \NW\cdot D^3$, this is the actual convergence of the MPS using $D$ from simulation ($\timeNESS$ is still from the exact evolution operator). The solid circle points are from the estimated costs using a scaled version of Eq.~\eqref{eq:EstCompCost}, $c \cdot \NW \cdot \timeNESS \cdot 2^{3\cdot \OSEE}$. We take $c=10^6$ as a uniform factor to match where the given $D$ is sufficient to achieve convergence at a set $\NW$. 
    The error of the current in MPS calculation is estimated by an average $\sig$, as defined in Eq.~\eqref{eq:CurrErr}, over a time window of duration $20/\omega_0$ for CR or 2 to 5 cycles for ARC.
    Time evolution uses an integration step of $0.125/\omega_0$. The refresh is applied in a single step rather than via time evolution.  
    }
    \label{fig:9}
\end{figure}

Tensor networks are a common choice in studying low dimensional quantum systems. Those methods provide a scalable and accurate tool to target models with many--body interactions~\cite{orus_tensor_2019,verstraete_matrix_2008,ran_tensor_2020}. The crux of the methods, though, is that there needs to be limited entanglement so that the simulation can target small dimensional subspaces of the total Hilbert space. We will employ an implementation using one--dimensional tensor networks based on matrix product state (MPS)~\cite{verstraete_matrix_2008,ran_tensor_2020} and evolved using the time-dependent variational principle (TDVP)~\cite{haegeman_time-dependent_2011,haegeman_unifying_2016}. More specifically, we will employ for MPS the {\em mixed basis} that simulates the reservoirs in their eigenbasis with $\ql$ and $\qr$ modes paired in a way that makes the entanglement entropy scale logarithmically in time~\cite{rams_breaking_2020} (opposed to prohibitive linear growth in time~\cite{cazalilla_time-dependent_2002} that occurs in transport due to each scattering event creating partially entangled particle--hole pair~\cite{beenakker_electron-hole_2006,klich_quantum_2009}). We include more comments on the mixed basis in Sec.~\ref{sec:cost}.

This method was implemented for open system dynamics within ERAs~\cite{wojtowicz_open-system_2020} by vectorizing the density matrix~\cite{zwolak_mixed-state_2004} (so the local physical dimension is the square of the local Hilbert space dimension). It is important to note that the problem is only mapped to a 1D representation (in frequency or momentum space) for the state (the MPS). The reservoirs, though, can be any dimension so long a they are non--interacting. For TDVP, the Hamiltonian is written as matrix product operator (MPO). The geometry of the model and the mixed basis yields a very compact Hamiltonian MPO, i.e., with small virtual bond dimension.  Refresh operators can be applied directly on the density matrix after integrating over time, i.e., no time evolution is required. We otherwise follow the approach of Ref.~\cite{wojtowicz_open-system_2020}.

It is important to emphasize that error--cost scaling and convergence with $D$ are both intertwined and heavily dependent on the tensor network implementation (and geometry and other aspects of the physical problem). For instance, using the same implementation but ordering the reservoir modes to retain the spatial $\ql - \qs - \qr$ structure, the cost estimate, Eq.~\eqref{eq:EstCompCost}, still applies. However, $\OSEE$ grows linearly in time and thus the cost would be many orders of magnitude larger than for our implementation. Put more bluntly, when working in the mixed basis, $D \sim t \sim \NW$ due to the logarithmic growth of the (operator) entanglement entropy, giving a total cost growing polynomially. Retaining the spatial structure, however, gives $D \sim 2^{\varsigma t} \sim  2^{\varsigma^\p \NW}$ with $\varsigma$ and $\varsigma^\p$ being some positive constant. Similar considerations apply to density matrix renormalization group methods or other MPS implementations that simulate one--dimensional reservoirs with only nearest--neighbor hopping~\cite{cazalilla_time-dependent_2002,zwolak_mixed-state_2004,gobert_real-time_2005,schneider_conductance_2006,al-hassanieh_adaptive_2006,dias_da_silva_transport_2008, heidrich-meisner_real-time_2009,branschadel_conductance_2010,chien_interaction-induced_2013,gruss_energy-resolved_2018}. Other implementations, though, may not have the same scaling as in Eq.~\eqref{eq:EstCompCostMPS}, and thus will not use the same approximate form, Eq.~\eqref{eq:EstCompCost}, in terms of the operator entanglement entropy.

During the ARC process, there are two alternating processes. The step involving unitary time evolution results in the spreading of correlations and, as a result, OSEE growth. On the other hand, during the refresh step dissipation breaks some entangled pairs, making the density matrix closer to the thermal state. The thermal state is a product state with OSEE equal to zero and with trivial physical dimension $D=1$. 

Figure~\ref{fig:9}(a) shows how the current error converges versus $\CC^\mathrm{MPS}$ (that employs $D$) for CR and $\gamma\tau=1$ at two different reservoir sizes. Truncation introduces a non--trivial level of error. This error decreases as we take $D\rightarrow\infty$ but is non--monotonic. For the CR regime, convergence is more well behaved than for ARC at $\gamma\tau=1$, which is somewhat expected given the lower level of $\OSEE$ in CR. PR has an even higher level of entropy and has higher cost in our tensor network implementation requiring an effort comparable to simulating closed systems but using a vectorized density matrix that squares the cost contribution from $D$. The plot also shows the full cost estimate scaled by a uniform factor (the same for all $\gamma\tau$ and $\NW$), showing that the estimates from non--interacting simulations (particularly, $D \sim 2^{\OSEE}$) are reasonable. Figure~\ref{fig:9}(b) shows the results for high temperature, which displays similar behavior but with convergence moderately improved and supporting the same conclusion. 

\section{Conclusions}
\label{sec:conclusions}

We introduced the {\it accumulative reservoir construction} (ARC) that provides a bridge between continuous relaxation (CR) and periodic refresh (PR) extended reservoir approaches. The ARC considers a periodic evolution alternating between coherent and dissipative dynamics. The duration of the coherent evolution, the strength of the relaxation, and the reservoir size yield a three parameter family that goes to CR and PR in two opposing limits. At fixed reservoir size, the phase diagram (drawn from the current) reflects aspects of Kramers' turnover, with a Zeno--like regime on one side going through an intermediate, moderate relaxation regime before the transition into an underdamped regime, as we characterized. {Due to stroboscopic sampling, PR and PR--like regimes manifest underdamped behavior differently than CR and CR--like regimes. Yet, the underlying physics is the same}

This turnover has been extensively investigated for CR for quantum electron transport~\cite{gruss_landauers_2016,elenewski_communication_2017,gruss_communication_2017,wojtowicz_open-system_2020, wojtowicz_dual_2021,elenewski_performance_2021} and classical thermal transport~\cite{velizhanin_crossover_2015,chien_thermal_2017,chien_topological_2018}. Examining simulation techniques across their full parameter space, rather than fixing parameters (e.g., setting $\gamma$ for CR and $\tau$ for PR to functions of $\NW$), elucidates the underlying machinery and assesses different facets of their behavior. For CR, this has enabled the derivation of the bound {$\beta\gamma \le 1$}~\cite{gruss_landauers_2016,elenewski_communication_2017}, which also sets a minimum reservoir size, and the discovery of anomalous regimes~\cite{gruss_landauers_2016,wojtowicz_dual_2021} that constrain how to employ the techniques, especially with modest computational resources~\cite{elenewski_performance_2021}. ARC makes this readily apparent. ERA--based simulation techniques have a complex interplay of timescales and phenomena. For PR, in particular, we have seen the role of $\tau^\star$ (the rise time), $\TS$ (the transit time across the system), and $\TT$ in determining behavior and performance. Just like CR, there is reservoir size set by the thermal correlation length, which serves as a threshold for accurate simulations. However, there are other contributions to the errors in both cases, of which a full exploration is necessary. Our approach---the scanning of $\tau$ for PR, in particular---illuminates the role of various timescales. 

The intermediate regime is the target for simulations, as it can provide accurate approximations to the steady--state current and NESS in the presence of continuum, macroscopic reservoirs. In other words, it is the physical regime of the ARC--class of techniques. It can be identified as a plateau region of almost-constant current and system density matrix.
When going from CR to PR in this physical regime, there is a ``cut'' (complete with discontinuous quantities, such as the entanglement entropy, when following optimal parameters) when transitioning between quasicontinuous and refresh--like physics. Away from the cut, the CR-- and PR--like regimes, have well--defined error--cost scaling, with the latter having approximately twice the error decay rate as the former. This more rapid decay, though, comes at a cost of a higher overall prefactor due to a larger operator entanglement entropy. 

The best error--cost scaling---the best performance---for our example occurs in this transition regime (i.e., the intermediate ARC regime), right after the cut on the refresh side but well before reaching parameters that approximate the full PR. As with PR, this scaling comes at a cost (albeit, lower) in the entanglement entropy. This will be challenging to converge, as is readily apparent in the tensor network simulations. Higher temperatures reduce this barrier, bringing PR and other ARC regimes into more computationally accessible regions. 

{We expect many of the results for the resonant--level model to generalize to other models, including many--body problems. For continuous relaxation, the existence of overdamped, $1/\gamma$, behavior of the current has been proven to exist for all models (regardless of reservoir spectral function and impurity system, including in the presence of interactions)~\cite{gruss_landauers_2016}. The underdamped---relaxation--limited---regime also has been proven for all proportionally coupled, non--interacting impurities~\cite{zwolak_analytic_2020}. Given the intuitive nature of this regime---that the relaxation limits the rate of particle input/output---it likely holds for all impurities, including those with many--body interactions. For PR physics, we also expect the same based on intuitive arguments. The region of the phase diagram that should change the most is the ``cut'' between CR-- and PR--like physics. However, we only expect this to change quantitatively, not qualitatively. The behavior of entanglement entropy should also be similar and thus the trade--off of computational cost and accuracy should still hold. A full exposition of this is left for future work.}

While we have focused on characterizing ARC, especially its phase diagram and performance, the discrete nature of the ARC approach makes it suitable for experimental realization, such as in ultra--cold atoms and ions~\cite{muller_engineered_2012}. The flexibility of these experimental approaches in controlling interactions, as well as spectral functions, temperatures, and other parameters, together with ARC may provide a route to accurate physical simulators of open, many--body quantum systems. Unlike (classical) computational simulation, entanglement generated during the dynamics is not an explicit cost. Thus, ARC (and PR, in particular) may provide a promising framework to include reservoirs~\cite{harrington_engineered_2022}, as well as generate novel dynamical states of matter.

\section*{Acknowledgements}
 This research has been supported by National Science Centre, Poland, under project 2020/38/E/ST3/00150 (G.W. and M.M.R.).
 A.P. acknowledges funding from the European Research Council (ERC) under the European Unions Horizon 2020 research and innovation program (Grant Agreement No. 758403). A.P. also acknowledges funding from the Danish National Research Foundation through the Center of Excellence ``CCQ'' (Grant agreement no.: DNRF156)

\section*{Appendix}

\appendix

\renewcommand{\theequation}{S\arabic{equation}}
\setcounter{equation}{0}
\renewcommand{\thefigure}{S\arabic{figure}}
\setcounter{figure}{0}

In the first part of the appendix, we will provide further discussion on Kramers' turnover for CR and PR. Next we will discuss ARC and its phase diagram when changing reservoir and system $\qs$ sizes. In the ARC section, we will further review our averaging procedure to obtain $\asig$ and scaling relations at optimal point. Finally, we describe the procedure to obtain the operator entanglement entropy for non--interacting fermions. 

\section{Kramers' turnover and optimal $\action$ for CR}\label{si:cr}

\begin{figure}[h!]
    \centering
    \includegraphics[width=\columnwidth]{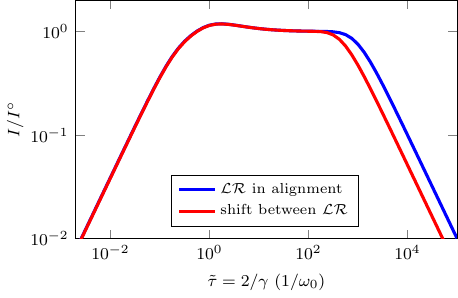}
    \caption{{\bf CR: Kramers' turnover and estimating the optimal $\gamma$.} The relaxation rate significantly influences the steady--state current. There are always three distinct regimes of behavior~\cite{gruss_landauers_2016}. Examining $I/\Io$ versus $\action$, these are seen here as the current initially increasing linearly with $\action \sim \gamma^{-1}$ (overdamped), plateauing (physical), and then decreasing as $1/\action \sim \gamma$ (relaxation--limited). There also can be anomalous regimes depending on the particular model~\cite{wojtowicz_dual_2021}, for which we see here a small Markovian anomaly on the left hand side of the plateau. In order to estimate the optimal $\gamma$ (or $\action$), one can shift the alignment of modes in $\ql$ and $\qr$, which removes virtual tunneling between them. The $\ql\qr$ in alignment and shifted intersection provides a good estimate of the physical current for given $\NW$. The heuristic approach avoids unphysical features by targeting a $\gamma$ that is sufficiently large that the discreteness of the modes is not influencing the transport but allows coherence to develop~\cite{elenewski_performance_2021}. The plot uses $\NW=512$ sites per reservoir, temperature $\beta\omega_0=40$, and bias $\mu=\omega_0/2$.}
    \label{fig:s1}
\end{figure}

The CR limit of ARC is when $\tau\rightarrow 0$, which gives the equation of motion in Eq.~\eqref{eq:evolution_cr}. The steady--state current is strongly influenced by the relaxation rate $\gamma$~\cite{velizhanin_crossover_2015,gruss_landauers_2016,elenewski_communication_2017,wojtowicz_open-system_2020,wojtowicz_dual_2021}. 
Figure~\ref{fig:s1} shows this influence versus $\action=2/\gamma$ (for CR), with the current going through the three regimes (from left to right, overdamped to physical to relaxation--limited, see Sec.~\ref{sec:cr}). We stress that in prior works these Kramer turnover plots showed the current versus $\gamma$ rather than action $\action$, so all regimes are the mirror reflection.

As a simulation approach, one wants the optimal $\gamma$ to target the physical regime and be as far as possible from anomalous behavior~\cite{elenewski_performance_2021,wojtowicz_dual_2021}. While generally one expects $\gamma$ to be comparable to the level spacing to optimize the representation of the spectral function, anomalous behavior in certain models can make this a poor choice for the modest number of modes simulation techniques employ.

In Ref.~\cite{elenewski_performance_2021}, we proposed a simple heuristic approach to estimate the optimal $\gamma$ in a way that it applies not only to non--interacting systems but also many--body cases. Markovian relaxation of finite reservoirs can yield two anomalies, a Markovian anomaly (broadening of the occupied density of states rather than occupying the broadened density of states) and a virtual anomaly (alignment of reservoir modes opens unphysical channels of conductance due to the finite size of the reservoirs)~\cite{wojtowicz_dual_2021}. Shifting the reservoir modes so that $\ql$ and $\qr$ modes do not align removes the latter. Finding where the Kramers' turnover with a shift converges to the turnover with $\ql\qr$ in alignment yields a $\gamma$ where the current and other properties are stable with respect to small perturbations of the reservoir mode placement, and the current {\em should be} independent of such details. Here, this convergence occurs as a crossing of the two curves. This approach avoids the unphysical features, including both under-- and over--damping, but also the Markovian and virtual anomalies. It thus provides a practical, effective approach to take the continuum limit that recognizes we are always at finite $\NW$ and $\gamma$.

\section{Physics and convergence for PR}
\label{si:pr}

\begin{figure}[t!]
    \centering
    \includegraphics[width=\columnwidth]{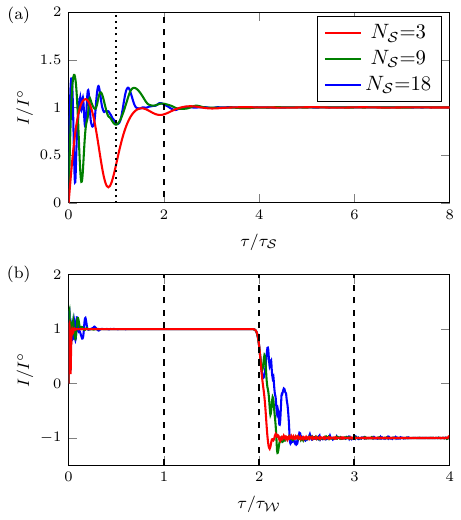}
    \caption{{\bf Typical regimes for PR.} (a) The stroboscopic steady--state current versus $\tau/\TS$ for three system sizes (giving three different $\TS$). For very small $\tau$, the current rises to the steady--state level. If the refresh period $\tau$ is comparable to any timescale in $\ql\qs\qr$, then coherent oscillations become apparent. For $\tau < \TS$, for $\TS$ the $\qs$ timescale, we observe features in the steady--state current related to evolution inside $\qs$, and distinct features for multiples of $\TS$ that match refresh time with wavefront traversing the $\qs$ from one side to the other. The features become less apparent as $\tau$ increases, eventually becoming negligible. The vertical dotted and dashed lines in (a) mark $\TS$ and $2\TS$, respectively. 
    (b) The stroboscopic steady--state current versus $\tau/\TW$ for the same set of system sizes. The current remains stable even after reaching the reservoir timescale $\TW$. It is only when the wave front returns to the interface at $2\TW$ that the current reverses, for which there will be further reversals at integer multiples of $2\TW$ (i.e., $4\TW$, etc.). The vertical dashed lines in (b) mark multiples of $\TW$. After reversal, the timescale $\TS$ determine when the current stabilizes. Unlike in the main text, $\qs$ is a uniform lattice of $\NS$ sites with $\onS=0$ and with nearest--neighbor hopping $v_\qs=\omega_0/2$. The data are obtained for reservoirs with $\NW=256$ modes each, temperature $\beta\omega_0=40$, and symmetrically applied bias $\mu=\omega_0/2$.}
    \label{fig:s2}
\end{figure}

Just like CR, PR also goes through a turnover. Since examining this turnover is part of this work, we describe it in some detail here, similar to how we summarize prior work on CR in Sec.~\ref{sec:cr}. At a fixed $\NW$, the (stroboscopic NESS) current exhibits the following regimes as $\tau$ varies:
\begin{itemize}
    \item {\em Small--$\tau$} or {\em Zeno} regime. In this regime, the refresh time is too short, suppressing coherence between the extended reservoir and $\qs$ that is necessary for a current to flow. This effectively decouples the extended reservoirs, resulting in a current decreasing with $\tau$. The situation here is the Zeno effect. It will occur when the refresh frequency is faster than the timescales for flow from/to $\ql$ and $\qr$. The rise time, $\tau^\star = \pi/\qw$, for an electronic reservoir of bandwidth $\qw$, see Ref.~\onlinecite{zwolak_communication_2018}, plays this role for our example. During this time, the interfacial currents increase linearly after the reservoirs are put into contact with the system. The refresh cuts off this increase. This regime is analogous to the large--$\gamma$ regime for CR, as seen in Fig.~\ref{fig:4}(b).
    
    \item {\em Moderately--small--$\tau$} or {\em overly--coherent--$\qs$} regime. This regime has no analog in CR. Here, particles flow into the system, but their coherence with the reservoir is destroyed by the refresh before coherence between all parts of $\ql\qs\qr$  is sufficient to have natural current flow. The regime emerges from interplay between multiple timescales and phenomena. For smaller $\tau$, current oscillations due to Gibbs phenomenon~\cite{zwolak_communication_2018} at the system--reservoir interface influence the NESS current as the refresh can come at points near the maximum or minimum of the oscillations. Simultaneously, this leaves excess moving particles within the $\qs$ that oscillate---i.e., remnants of the current that are coherent in $\qs$ but incoherent with $\ql\qr$. As a result, the current will vary depending on the interplay between $\qs$ timescales and $\tau$. In particular, there will be a transit time across $\qs$ of $\TS=\NS/2v_\qs$ set by the Lieb--Robinson bound. This creates oscillations in the NESS current as $\tau$ varies, as seen in Fig.~\ref{fig:4}(b), with a distinct minimum at $\TS$. At larger multiples of $\TS$, there are also oscillations, but their visibility will be model dependent. Additional timescales within $\ql\qs\qr$ could potentially play a role for more complex models. When moving away from the PR into ARC, this regime defines the cut between PR--like and CR--like behavior, as we discuss extensively in the main text.
    
    \item {\em Intermediate--$\tau$} or {\em physical} regime. In this regime, the refresh time is sufficient to support the formation of the intrinsic, physical current. That is, the coherence across $\ql\qs\qr$ is global. A plateau in the current develops and increases in size with $\NW$. This plateau is a reflection of the quasisteady state that forms within microcanonical---closed $\ql\qs\qr$--- approaches~\cite{ventra_transport_2004,bushong_approach_2005,chien_landauer_2014}, of which PR is most similar. It is in this regime that the current---typically the most important observable---accurately captures the true NESS current. As well, the approximate NESS approaches $\rhoo$. Extending the plateau is the limit $\NW \to \infty$.

    \item {\it Large--$\tau$}  or {\it overly coherent $\ql\qs\qr$} regime. In this regime, the particle current at the stroboscopic points is influenced by the timescale for coherent flow across the whole of $\ql\qs\qr$. The time between refresh events is too long compared to this transit time. The wavefront of particle flow scatters off the finite boundaries of $\ql\qs\qr$. The complete reversal of the current happens at $2\TW$, given by a Lieb--Robinson bound for the reservoir, when the reflected wavefront comes back to the interface. This occurs for $\tau$ larger than $\TW$, although the system transit time will also play a role in the current reversal, as seen in Fig.~\ref{fig:s2}. This is intuitively clear since the wavefront needs to travel across the system as well. The overly--coherent oscillations that occur in this regime also occur for the relaxation--limited regime of CR, but there the relaxation strength ultimately limits the current because a global NESS develops in $\ql\qs\qr$. {In other words, the PR and PR--like current displays a glaring over coherence rather than linear dependence on $1/\tau$ (i.e., something more directly analogous to CR) because one is using stroboscopic points to provide the NESS. If one employed the average current, then the current would oscillate about $1/\tau$, still giving signatures of over coherence but also reflecting that the frequency of refresh is the rate--limiting process for the overall current.}
\end{itemize}

Figure~\ref{fig:s2} shows how $\tau$ influences the stroboscopic NESS current. Figure~\ref{fig:s2}(a) shows three different lengths of $\qs$, giving distinct transit timescales, $\TS$. For larger $\TS$, the overly--coherent $\qs$ regime extends to larger $\tau$. The point $\tau/\TS = 1$ shows a qualitative change in the curves, with strong oscillations prior to this point. Figure~\ref{fig:s2}(b) shows that when the refresh time equals $2\TW$, the wavefront has scattered off the boundary of $\ql\qs\qr$ and arrived back at the interface with $\qs$. This results in reversal of the current for all $\qs$. The system's timescale also plays a role in this reversal, showing up as oscillations before the current fully stabilizes in the reverse direction.

\begin{figure}
    \centering
    \includegraphics[width=\columnwidth]{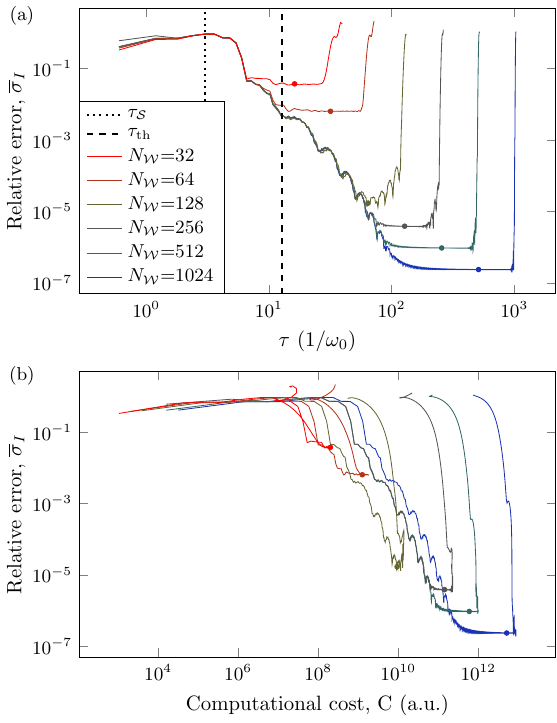}
    \caption{{\bf PR errors for increasing $\bm{\NW}$.} (a) The curves show the relative error versus $\tau$ for many different $\NW$. As $\NW$ increases, the plateau for physical behavior increases. As well, more reservoir modes increase the accuracy of the current. The vertical dotted line shows the impurity timescale $\TS$ and the vertical dashed line shows thermal time $\TT=\NT/2\omega_0$ for a reservoir with thermal relaxation time $\beta\omega_0=40$. (b) The curves show the same results as in (a) but versus the computational cost, Eq.~\eqref{eq:EstCompCost}.}
    \label{fig:s3}
\end{figure}

Moreover, when increasing the size of reservoirs, the plateau increases between overly--coherent $\qs$ at moderate (and $\NW$ independent) $\tau$ and overly--coherent $\ql\qs\qr$ at $\tau=2\TW$. Figure~\ref{fig:s3} shows this increase in plateau size. Additionally, the larger $\NW$ increases the accuracy of the current when compared to the continuum limit. This results in a plateau reaching lower $\asig$ error with increasing $\NW$, as is visible in Fig.~\ref{fig:s3}.

\section{ARC}
\label{si:arc}

\subsection{Kramers' turnover analog}
\label{si:arc_kramers}

\begin{figure}
    \centering
    \includegraphics[width=\columnwidth]{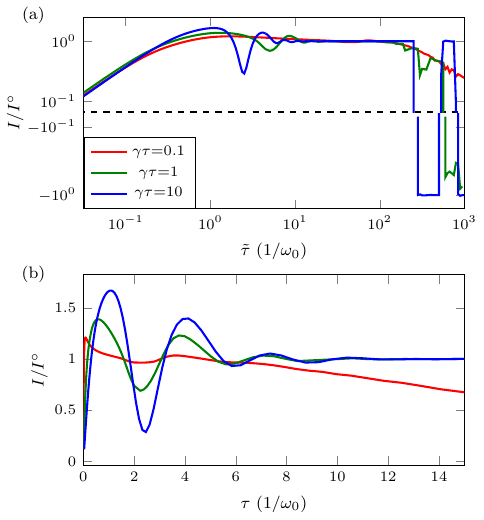}
    \caption{{\bf Steady--state current for ARC.} (a) The curves show the stroboscopic steady--state current for ARC versus $\action$ for a few $\gamma\tau$. All the parameter regimes show a Kramers'--like increase and then plateau for small to moderate $\action$. However, for larger $\action$ and $\gamma\tau$ the curves escape the CR--like phase and coherent oscillations become apparent. {For very large $\action$, all curves enter the underdamped regime, which can be either directly similar to a Kramers' rate--limiting process (i.e., a linear dependence on $\gamma$) or reflect the stroboscopic, PR--like current, which breaks down and eventually reverses.} For $\gamma\tau = 1$, however, the underdamped regime combines these two features and the current goes down with $\action$ but still undergoes some rapid oscillation and eventually reverses when the wavefront returns back to the interface. (b) The curves show ARC with the same three $\gamma\tau$ versus $\tau$ instead of $\action$ (the $\gamma$ is changing with $\tau$ since $\gamma\tau$ is fixed). The coherent oscillations for the two larger values of $\gamma\tau$ are seen to occur with the same frequency. Both plots take data from Fig.~\ref{fig:4}(b).}
    \label{fig:s4}
\end{figure}

ARC displays behavior that is both CR--like and PR--like depending on the point on the parameter space. Generally speaking the separation between these two regimes happens along a line of constant $\tau$ marking the smallest $\ql\qs\qr$ timescale to appear. Figure~\ref{fig:s4}(a) shows the Kramers' turnover analog for different ARC cases. When $\gamma\tau$ line goes fully inside CR--like regime, the behavior is (by definition) directly analogous to Fig.~\ref{fig:s1}. As $\gamma\tau$ increases, the curve starts to register coherent oscillations that appear on top of plateau. For low and moderate $\action$, the plot registers overly--coherent $\qs$, and then enters the physical PR--like regime which is the most natural scenario for the steady--state. Eventually, coherent evolution leads to the complete reversal of the current flow when entering the overly--coherent $\ql\qs\qr$ regime. Figure~\ref{fig:s4}(b) shows the same data but plotted versus $\tau$ instead of $\action$. This helps to see the emergence of oscillations as a result of coherent evolution. {The period of coherent oscillations is roughly constant with refresh strength. This is due to the fact that the (partial) refresh incoherently adds a stationary component to the reservoir's state (e.g., the thermal equilibrium state when not connected to the junction system). The component of the state carrying the wavefront will have it continue to propagate at the same speed (dictated by the Hamiltonian and initial packet). Even in the CR and CR--like regime, the wavefront propagates at the same velocity, but, since the  $\ql\qs\qr$ state goes into a global NESS, the wavefront completely dissipates. Its transversal, though, does influence the correlations present in the NESS. }

\subsection{Phase diagram in $\gamma-\tau$ space}
\label{si:arc_phasediagram2}

\begin{figure}[h!]
    \centering
    \includegraphics[width=\columnwidth]{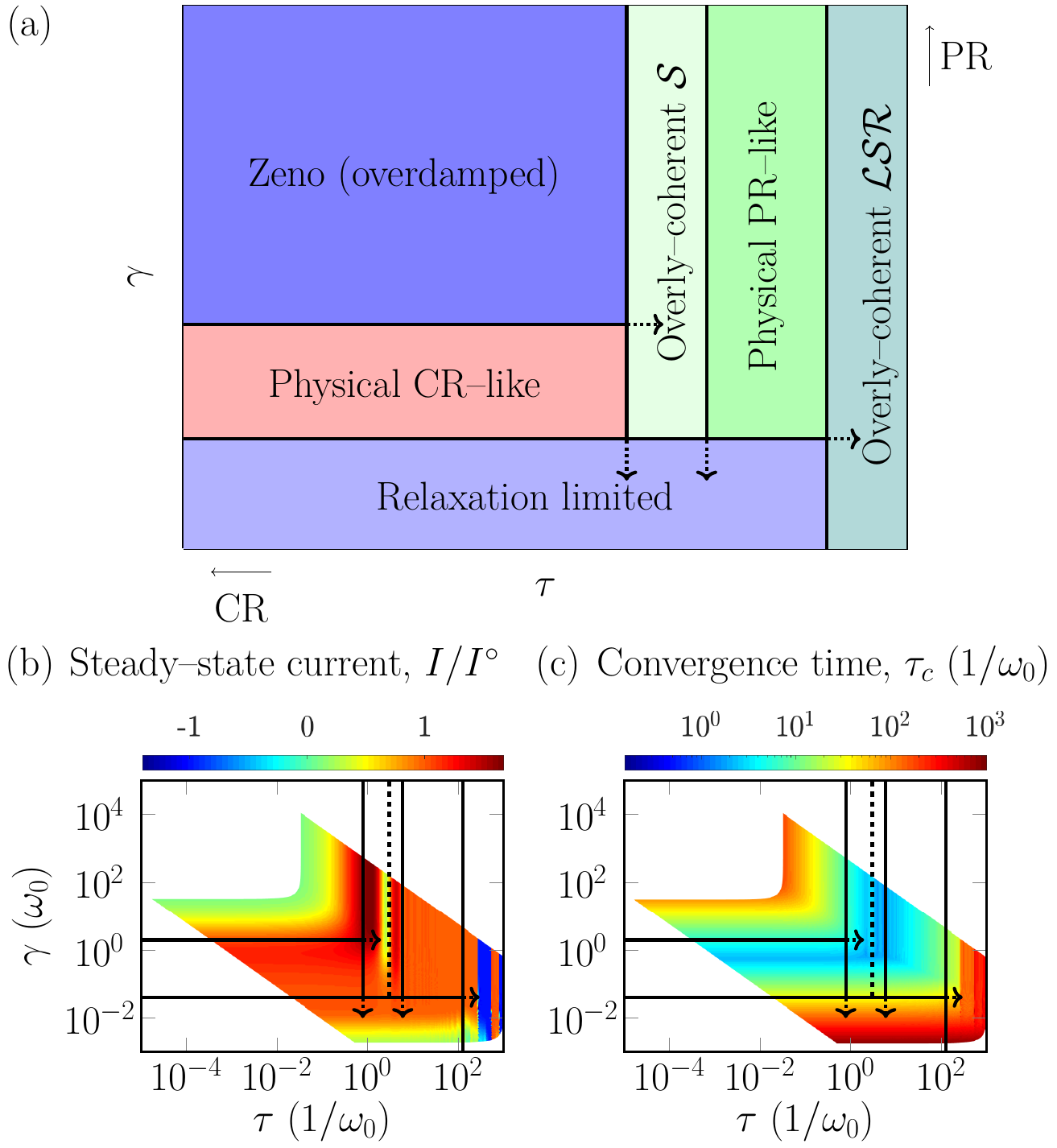}
    \caption{{\bf ARC phase diagram in $\bm{\gamma}$--$\bm{\tau}$ space.} (a) The abstract phase diagram for ARC. The lines of constant $\tau$ and constant $\gamma$ are now vertical and horizontal, respectively. The PR and CR limits are now on the top and left, respectively. This provides a different perspective on the phase diagram in Fig.~\ref{fig:4}. (b,c) The steady--state current and convergence time both on the $\gamma$--$\tau$ space. The data and lines of constant $\tau$ and $\gamma$ are at the same values as in Fig.~\ref{fig:4}.}
    \label{fig:s5}
\end{figure}

The steady--state induced by the ARC procedure results in the phase diagram Fig.~\ref{fig:4} which has a nice and regular structure when drawing it on $\gamma\tau-\action$ plane. However, the phase diagram boundaries are a consequence of $\gamma$ and $\tau$ taken separately. We can see them more vividly after changing the parameterization of the phase diagram to $\gamma-\tau$. In Fig.~\ref{fig:s5} the regimes are separated by vertical lines of constant $\tau$ and horizontal lines of constant $\gamma$. One can still find the limit of CR and PR, which now share a corner with the Zeno regime. Although, this point of view can be useful, it is the $\gamma\tau-\action$ parameterization which is more convenient and physically motivated. In Fig.~\ref{fig:4}, the phase diagram separates what is lost due to refresh, $\sim\gamma\tau$, from what remains, $\sim\action$. 

\subsection{Phase diagram for growing $\qs$ and $\ql\qr$}
\label{si:arc_growNSNW}

\begin{figure}[h!]
    \centering
    \includegraphics[width=\columnwidth]{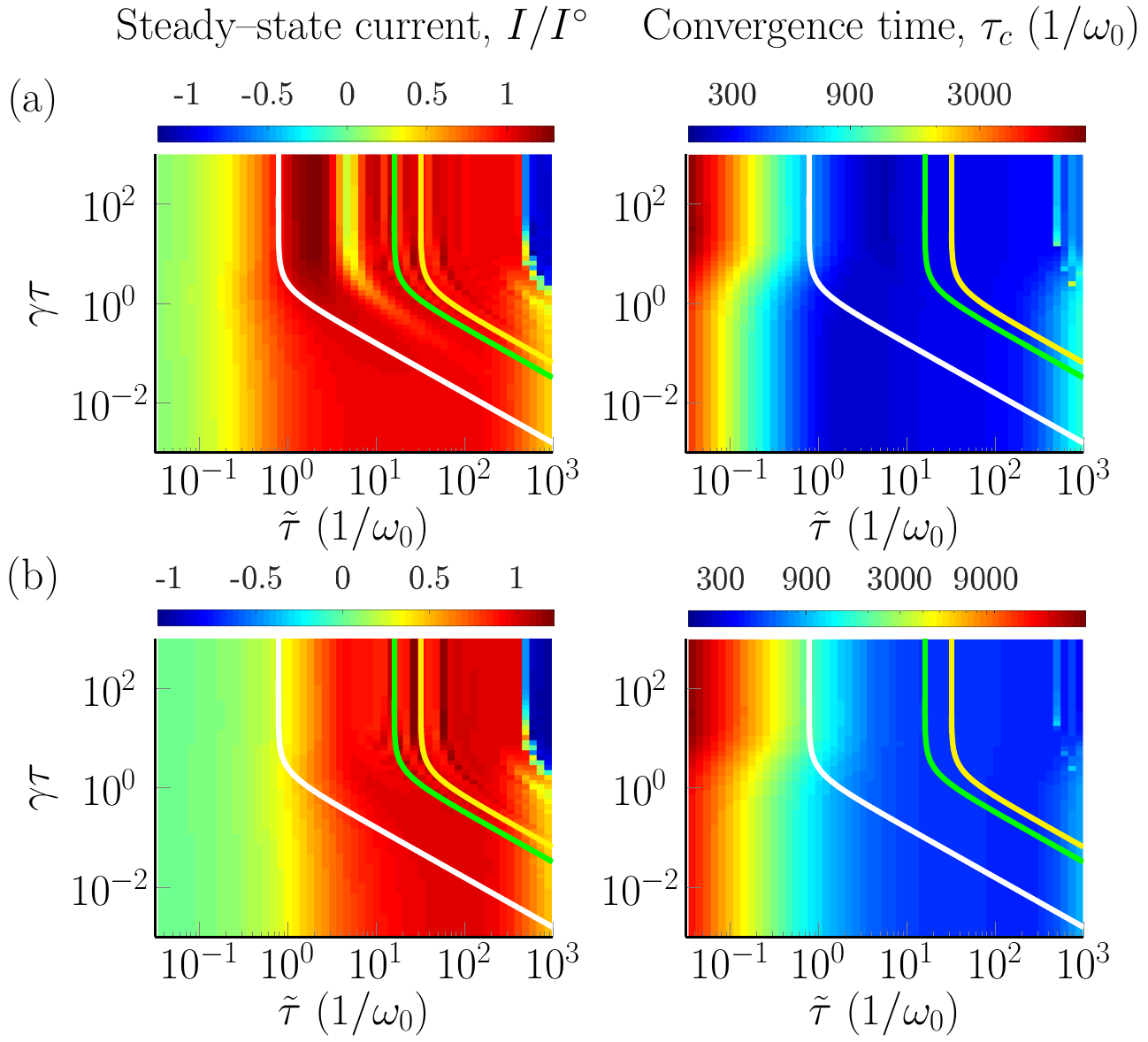}
    \caption{{\bf Phase diagram for $\bm{\NS=16}$.} The lines demarcate the boundary and some features of the overly--coherent $\qs$ regime. Unlike the main text, $\qs$ is a uniform lattice with onsite frequencies of zero, a constant nearest--neighbor coupling $v_\qs=\omega_0/2$, and system--reservoir coupling of (a) $\omega_0$ and (b) $\omega_0/2$. The line of fixed $\tau$ equal to $\tau^\star = \pi/\qw$ is plotted in white, $\TS$ plotted in green, and $2\TS$ plotted in yellow. The $\qs$ timescale is $\TS=\NS/2 v_\qs$. Both examples use $\NW=256$ sites per reservoir, temperature $\beta\omega_0=40$, and bias $\mu=\omega_0/2$.}
    \label{fig:s6}
\end{figure}

The phase diagram with distinct regimes as presented in Fig.~\ref{fig:4}(a) is universal for all $\qs$ sizes. In Fig.~\ref{fig:s6}(a), we show the phase diagram for uniform $\qs$ with no on-site energies and nearest--neighbor hopping $v_\qs=\omega_0/2$ coupled to reservoirs with amplitude $\omega_0$. The plot keeps the same regimes as for Fig.~\ref{fig:4}(b) but has wider regime of overly--coherent $\qs$ since the $\qs$ timescale is now larger. In Fig.~\ref{fig:s6}(b), we show the same case as in Fig.~\ref{fig:s6}(a) but now the coupling to reservoirs is $\omega_0/2$. In  Fig.~\ref{fig:s6}(b), the  phase diagram remains very close to  Fig.~\ref{fig:s6}(a) with the same upper bound for the Zeno regime. 

\begin{figure*}[t!]
    \centering
    \includegraphics[width=\textwidth]{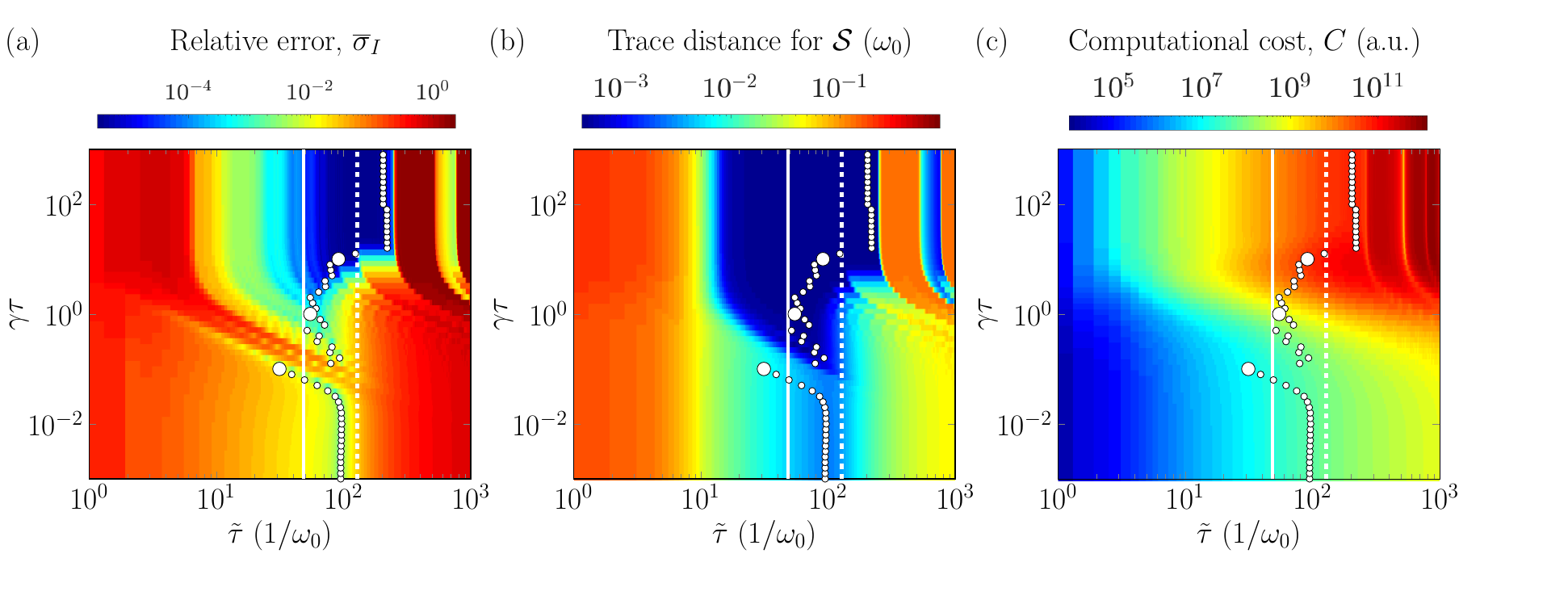}
    \caption{{\bf Convergence and computational cost of ARC.} The three panels show the (a) current error, Eq.~\eqref{eq:CurrErr}, (b) the correlation matrix error, Eq.~\eqref{eq:CorrErr}, and (c) the estimated computational cost, Eq.~\eqref{eq:EstCompCost}, for our example system and for $\NW = 256$, $\beta\omega_0=40$, and $\mu=\omega_0/2$. The two error plots have many of the same features as the phase diagram for ARC, Fig.~\ref{fig:4}, as we expect since the different regimes directly impact the error. Using the current error as the quantity to be optimized, the best choice of action for a fixed $\gamma\tau$ is marked by the white points in (a). These points are replicated in the plot showing the correlation matrix error (b), and estimated computational costs (c) to see where they fall for other quantities. 
    The whites dots for $\gamma\tau=0.1, \, 1$, and $10$ are made thicker to guide readers eye. Vertical solid line represents $\action$ equal to heuristic choice for CR, and vertical dashed line represents $\action=\TW$, which is a physical choice for PR. We employ these $\action$ for the CR and PR limits.}
    \label{fig:s7}
\end{figure*}

Some boundaries in the phase diagram, Fig.~\ref{fig:4}, will depend on $\NW$. The transition into the Zeno regime and the transition from CR--like to PR--like both depend on intrinsic characteristics of the continuum model but not on $\NW$ (except for some possible finite--size effects at really small $\NW$). When $\NW$ increases, the transition from the physical regimes to underdamped regimes is shifted to higher $\action$. This makes the physical regimes wider but otherwise maintains the structure of the phase diagram. The impact of $\NW$ on phase diagram can be seen by comparing Fig.~\ref{fig:s7} for $\NW=256$ to Fig.~\ref{fig:5} for $\NW=512$. As a consequence of growing $\NW$, the $\gamma\tau$ which initially had a CR--like physics at the optimum, and corresponding scaling relation, can transition to PR--like physics. For $\gamma\tau=0.1$, the optimal point for $\NW=256$ in the CR--like regime moves to the PR--like regime for $\NW=512$. 

\begin{figure}[h!]
    \centering
    \includegraphics[width=\columnwidth]{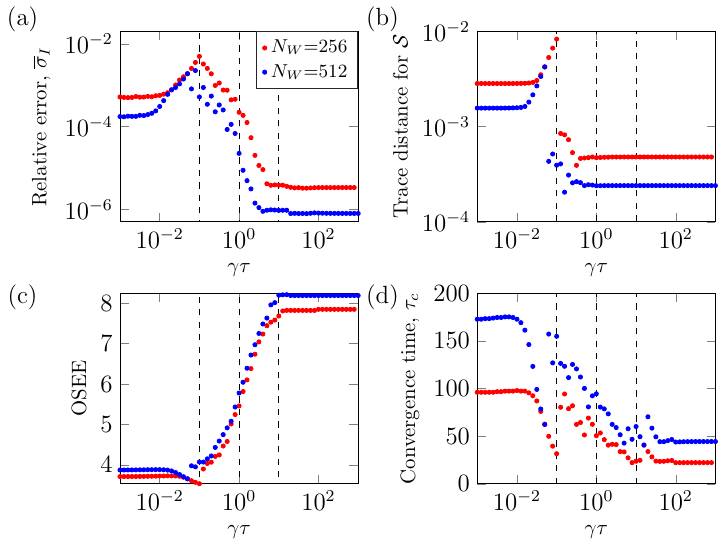}
    \caption{{\bf Optimal estimates for ARC.} The data take ARC optimal points  from Fig.~\ref{fig:s7} for $\NW=256$ and  from Fig.~\ref{fig:5} for $\NW=512$. From the white points we extract (a) relative error $\asig$, (b) error for $\qs$ correlation matrix, (c) operator space entanglement entropy for central cut, and (d) the convergence time to reach a steady state $\timeNESS$. 
    The data show the transition between optimal estimates falling into CR--like regime to these which are already in PR--like regime. Vertical dashed lines are included to guide readers eye and mark $\gamma\tau=0.1, \, 1$ and $10$, which are highlighted points in Fig.~\ref{fig:s7} and Fig.~\ref{fig:5}.}
    \label{fig:s8}
\end{figure}

One can also see the influence of the different parameter phases when examining quantities versus $\gamma\tau$. Figure~\ref{fig:s8} shows errors (in current and trace distance) and contributions to the cost (OSEE and convergence time) for the optimal points taken from Fig.~\ref{fig:s7} and Fig.~\ref{fig:5}. There is a clear distinction between points in the physical CR--like regime and those in the physical PR--like regime, as well as intermediate ARC for the cases in between. For the physical CR--like regime, the error of the current saturates at a value approximately $100\times$ larger than for the physical PR--like regime. The same happens for the trace distance, which has an approximately $10\times$ larger value (i.e., error) for physical CR--like regime. Better accuracy comes from longer coherent evolution in the physical PR--like regime that  accumulates more non--Markovian character. 
However, it comes with a penalty: The OSEE increases by approximately 2 times. This makes the bond dimension increase quadratically when going from physical CR-- to PR--like ARC. In fact, the increase of entropy dominates the overall computational cost in Eq.~\eqref{eq:EstCompCost}. 

For the intermediate $\gamma\tau$, there a discontinuity in the trace distance, OSEE, and convergence time. This is due to the optimal $\action$ jumping from the CR--like to the PR--like side of the phase diagram. Since $\asig$ is the quantity used to assign the optimal point (and it is a smooth function of $\gamma\tau$ and $\action$), there will be no discontinuity in it. The convergence and cost for intermediate ARC interpolates between physical CR-- and PR--like ARC.

\subsection{Averaging procedure to get $\asig$}
\label{si:arc_averaging}

\begin{figure}[h!]
    \centering
    \includegraphics[width=\columnwidth]{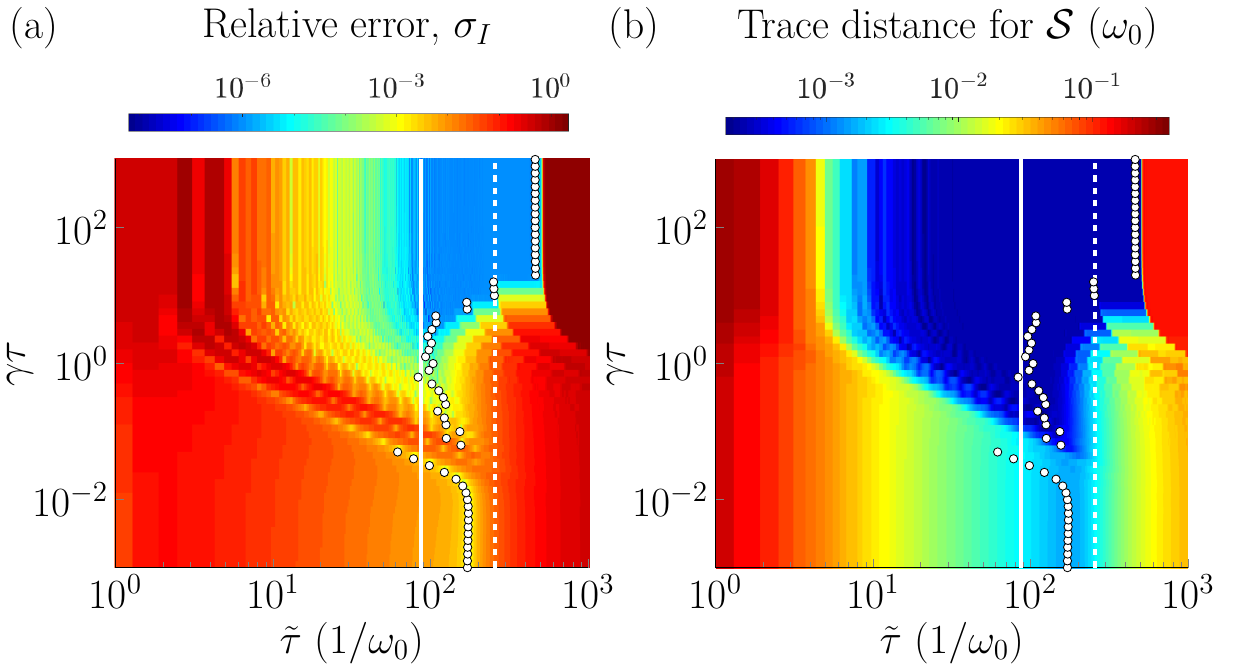}
    \caption{{\bf Relative error without the moving average.} The panels show the (a) current error, $\sig$, (b) the correlation matrix error, Eq.~\eqref{eq:CorrErr}, for  $\qs$ as in the main text. The reservoirs are  $\NW = 512$ sites each, temperature $\beta\omega_0=40$, and symmetric bias $\mu=\omega_0/2$. The plot takes the data from Fig.~\ref{fig:5} but does not apply moving average in Eq.\eqref{eq:sigmean} to $\sig$. White points are taken from Fig.~\ref{fig:5} and mark the position of lowest average error $\asig$ for each $\gamma\tau$. These plots demonstrate that the averaging procedure does not deform the phase diagram. They also show that the averaging itself does not bias the optimum. However, on the CR side, the optimum from $\asig$ tracks the accidental crossing. Those crossings are not fully visible on either the PR or CR side of the phase diagram due to the finite resolution of the heat map.}
    \label{fig:s9}
\end{figure}

As we discuss extensively in the main text, an  averaging procedure is taken in order to avoid assigning accidental crossings, where $I=\Io$, an error of zero. These crossings are due to the parameter sweep, rather than an accurate, physical representation of the problem. The moving average smooths out---regularizes---the error across these points, enabling the minimization procedure to choose an optimal $\action$---the minimum average error---at a fixed $\gamma\tau$, see Sec.~\ref{sec:convergence}. This provides an algorithmic approach to obtaining, e.g., the error--cost scaling. It is by no means, however, ideal, since the optimal point still can lie close to the crossings even if the error is more appropriately assessed. This is why for CR and PR we choose the well--defined heuristic points for setting the action instead. We compare their scaling relations to these we obtain with the $\asig$ minimization procedure in similar ARC regimes. Figure~\ref{fig:7} shows that this algorithmic approach is sound for delineating scaling regimes (i.e., exponents), but the error of CR--like behavior is still underestimated, as seen by the comparison with the CR limit. 

To visualize the influence of the moving average, Fig.~\ref{fig:s9} presents the data from Fig.~\ref{fig:5} without the moving average. Figure~\ref{fig:s9} has many accidental crossings with $\Io$, albeit they are not fully visible due to the finite resolution of the plot. These can lead to wrongly assigning the $\action$ for capturing the physical problem when taking $\sig$ directly. The trace distance in both Fig.~\ref{fig:5} and Fig.~\ref{fig:s9} does not have averaging, since regularizing the data was only necessary for determining the optimal $\action$ via the current. We further note that Fig.~\ref{fig:s9} demonstrates that the averaging procedure does not significantly distort or otherwise modify the phase diagram. It also shows, however, that the optimal points from $\asig$ are influenced by the accidental crossings, especially in the CR regime. For the full CR limit, though, we use a different procedure to choose the action, one that is not influenced by the accidental crossings, as we discuss extensively in the main text and in the first appendix. In the PR regime, the optimal points are pushed to the right on the plateau, which increases accuracy marginally but also increase the computational cost. This is not accounted when assessing the scaling for ARC. However, for the PR limit, we use the physical time scale, which shows the scaling is not significantly modified by using the optimal point. As we mention earlier, if one takes a $\tau$ smaller than the physical tau, $\TW$, in the PR regime, the cost decreases without a significant loss in accuracy [see Fig.~\ref{fig:s3}(b)].

\subsection{Scaling relations}
\label{si:arc_scaling}

In the main text, we discussed the error--cost scaling, providing some of the extracted values for the prefactors and exponents in Fig.~\ref{fig:7}. The fits corresponds to the accuracy and performance of the ARC at fixed $\gamma\tau$ taken at the optimal point. Here, we provide the remaining values, as well as contributions to these values. Table~\ref{tab:arc_low_T} and Table~\ref{tab:arc_high_T}, at low and high temperature, respectively, give the scaling relations for the individual quantities versus $\NW$ rather than cost. For the scaling, we employ the following relations:
\[ \label{eq:sigscale}
\asig=A N_\mathcal{W}^{-\nu}
\]
for the optimal $\asig$;
\[ \label{eq:SOscale}
2^{S_O}={A}{N_\mathcal{W}^\nu},
\]
for operator space entanglement entropy $\OSEE$; 
\[ \label{eq:tauscale}
\tau_c={A}{N_\mathcal{W}^\nu},
\]
for convergence time $\timeNESS$; and
\[ \label{eq:costscale}
\CC={A}{N_\mathcal{W}^\nu}.
\]
for the cost $\CC=\NW\timeNESS 2^{3 \OSEE}$. The error--cost scaling relation has the form in Eq.~\eqref{eq:ErrScaling}. The values in the error-cost scaling can be fitted directly, see Table~\ref{tab:arc_error_cost}, or reconstructed from scaling relations of its components. For instance, the $\nu$ from Eq.~\eqref{eq:costscale} is the sum of three times $\nu$ from Eq.~\eqref{eq:SOscale} plus $\nu$ from Eq.~\eqref{eq:tauscale} plus 1. These approaches give consistent results and fit errors. Thus, the delineation of the fit parameters into their components enables pinpointing how different factors influence the simulation. 

\begin{table}[h!]
\caption{{\bf Scaling relations for low temperature, $T=\omega_0/40$.} The scaling relations are for data in Fig.~\ref{fig:7}(a).}
\begin{tabular}{l|c|c|c|c}
\diagbox[width=0.14\columnwidth, height=1cm]{ARC}{$\nu$} & $\asig=A N_\mathcal{W}^{-\nu}$ & $2^{S_O}=A N_\mathcal{W}^\nu$ & $\tau_c=A N_\mathcal{W}^\nu$ & $\CC=A N_\mathcal{W}^\nu$\\ \hline
{PR}  &{2.03 $\pm$ 0.02} & {0.393 $\pm$ 0.001} &{0.83 $\pm$ 0.01} &{3.01 $\pm$ 0.09}\\\hline
\begin{tabular}{l}ARC \\$\gamma\tau=1$\end{tabular} & 3.4 $\pm$ 0.2 & 0.23 $\pm$ 0.01 & 0.66 $\pm$ 0.04 & 2.37 $\pm$ 0.07  \\\hline
CR & 0.91 $\pm$ 0.02 & 0.171 $\pm$ 0.003 & 0.94 $\pm$ 0.02 & 2.45 $\pm$ 0.03
\end{tabular}
\label{tab:arc_low_T}
\end{table}

\begin{table}[h!]
\caption{{\bf Scaling relations for high temperature, $T=\omega_0/2$.} The scaling relations are for data in Fig.~\ref{fig:7}(b).}
\begin{tabular}{l|c|c|c|c}
\diagbox[width=0.14\columnwidth, height=1cm]{ARC}{$\nu$} & $\asig=A N_\mathcal{W}^{-\nu}$ & $2^{S_O}=A N_\mathcal{W}^\nu$ & $\tau_c=A N_\mathcal{W}^\nu$ & $C=A N_\mathcal{W}^\nu$\\ \hline
{PR} &{1.96 $\pm$ 0.03} &{0.118 $\pm$ 0.001} &{0.82 $\pm$ 0.02} &{2.18 $\pm$ 0.02}\\\hline
\begin{tabular}{l}ARC \\$\gamma\tau=1$\end{tabular}  & 2.21 $\pm$ 0.04 & 0.079 $\pm$ 0.002 & 0.80 $\pm$ 0.01 & 2.04 $\pm$ 0.01\\\hline
CR & 0.87 $\pm$ 0.02 & 0.074 $\pm$ 0.005 & 0.94 $\pm$ 0.02 & 2.16 $\pm$ 0.03
\end{tabular}
\label{tab:arc_high_T}
\end{table}

\begin{table}[h!]
\caption{{\bf Scaling for error--cost.} The scaling relations correspond to dashed lines in Fig.~\ref{fig:7}[(a) and (b)].}
\begin{tabularx}{\columnwidth}{l|>{\centering\arraybackslash}X|>{\centering\arraybackslash}X}
\diagbox[width=0.14\columnwidth, height=1.1cm]{ARC}{$\ln{A}$,$\nu$} & $\beta\omega_0=40$ & $\beta\omega_0=2$\\ \hline
{PR}  & 4.2 $\pm$ 0.3, 0.67 $\pm$ 0.01 & 3.7 $\pm$ 0.4, 0.90 $\pm$ 0.02\\\hline
\begin{tabular}{l}ARC \\$\gamma\tau=1$\end{tabular}  & 21.7 $\pm$ 1.5, 1.44 $\pm$ 0.07 & 6.3 $\pm$ 0.4, 1.08 $\pm$ 0.02 \\\hline
CR & 2.14 $\pm$ 0.06, 0.372 $\pm$ 0.003 & 0.0 $\pm$ 0.2, 0.40 $\pm$ 0.01
\end{tabularx}
\label{tab:arc_error_cost}
\end{table}

\section{Operator-space entanglement entropy}\label{si:osee}
For completeness, in this section, we briefly summarize the calculation of OSEE in non--interacting simulations. To define OSEE, one treats a density matrix as a vector in a Hilbert space that is a tensor product of local operator spaces~\cite{prosen_operator_2007, dubail_entanglement_2017}. It amounts to mapping a density matrix into a pure state in the enlarged space containing auxiliary fermionic operators, deemed as adjoint fermions in Ref.~\cite{prosen_operator_2007} or a super-fermion representation in Fig.~\cite{dzhioev_super-fermion_2011}. Hence, for each site $m\in\ql\qs\qr$ we will have two fermionic species with annihilation operators $\f{c}{m}$ and $\uf{c}{m}$.
The normalized pure state is 
\[
\ket{\rho} \sim \rho \ket{I},
\]
where here we take $\ket{I} = \prod_{m\in\ql\qs\qr} (\fd{c}{m} - \ufd{c}{m}) \ket{0}$ as a product of maximally entangled singlets in each local space.

A Gaussian density matrix is fully determined by its correlation matrix $\cm$. In that case, a Gaussian $\ket{\rho}$ is described by the correlation matrix $\ucm$ in a block form
\[
\ucm_{mn} = 
\bra{\rho}\left[
\begin{array}{cc}
  \fd{c}{n} \f{c}{m}    &    \fd{c}{n} \uf{c}{m}  \\
 \ufd{c}{n} \f{c}{m}    &   \ufd{c}{n} \uf{c}{m} 
\end{array}
\right]\ket{\rho}.
\]
The OSEE is an entanglement entropy of bipartition of $\ket{\rho}$. It is calculated in a standard way~\cite{peschel_calculation_2003} from eigenvalues $\epsilon_i$ of a submatrix of $\ucm_{mn}$ with $m,n \in \mathcal{A}$, 

\[
\OSEE = - \sum_i {\epsilon_i \log_2{\epsilon_i}}  - \sum_i {(1- \epsilon_i) \log_2{(1-\epsilon_i)}}.
\]
In this work, we consider a bipartition in the {\it mixed} basis, which is pictorially represented in Fig.~\ref{fig:6}, and $\mathcal{A}$ is the group of modes defining bipartition.

Finally, to obtain $\ucm$, one rotates fermionic operators to a diagonal basis of $\cm$, where $\f{d}{i} = \sum_{m \in \ql\qr\qs} \qu^\dagger_{im} \f{c}{m}$  (similarly we will have $\uf{d}{i} = \sum_{m \in \ql\qr\qs} \qu^\dagger_{im} \uf{c}{m}$).
Here $\qu$ is a matrix of eigenvectors, and $\varepsilon_i$ are eigenvalues, of $\cm$. In that basis, a correlation matrix is block diagonal with
\[
\bra{\rho}\left[
\begin{array}{cc}
  \fd{d}{i} \f{d}{i}    &    \fd{d}{i} \uf{d}{i}  \\
 \ufd{d}{i} \f{d}{i}    &   \ufd{d}{i} \uf{d}{i} 
\end{array}
\right]\ket{\rho} = \left[
\begin{array}{cc}
  \alpha_i^2 &  \alpha_i \beta_i \\
  \alpha_i \beta_i &   \beta_i^2
\end{array}
\right],
\]
where $\alpha_i = \frac{\varepsilon_i}{\sqrt{(1-\varepsilon_i)^2 +  \varepsilon_i^2}}$, and $\beta_i = -\frac{1-\varepsilon_i}{\sqrt{(1-\varepsilon_i)^2 +  \varepsilon_i^2}}$. The correlation matrix $\ucm$ in the original basis of interest is obtained by rotating back to $\f{c}{m}$ and $\uf{c}{m}$.

\bibliography{main.bib}
 
\end{document}